\documentclass[prd,preprint,showpacs,superscriptaddress,nofootinbib,floatfix]{revtex4}
\usepackage{latexsym}
\usepackage{epsfig}

\def\br(#1,#2){\left\langle#1#2\right\rangle}
\def\sq(#1,#2){\left[#1#2\right]}
\def\s(#1,#2){s_{#1 #2}}
\def\t(#1,#2,#3){s_{#1 #2 #3}}

\begin{document}

\preprint{$
\begin{array}{l}
\mbox{ANL-HEP-PR-04-4}\\
\mbox{NSF-KITP-04-13}\\
\mbox{hep-ph/0403194}\\[5mm]
\end{array}
$}
\vspace*{2cm}

\title{Higgs Boson Production in 
Weak Boson Fusion at Next-to-Leading Order} 

\author{Edmond~L.~Berger}
\email[e-mail: ]{berger@anl.gov}
\affiliation{High Energy Physics Division, 
Argonne National Laboratory, Argonne, IL 60439}

\author{John Campbell}
\email[e-mail: ]{johnmc@hep.anl.gov}
\affiliation{High Energy Physics Division, Argonne National Laboratory,
Argonne, IL 60439}

\date{March 22, 2004}

\begin{abstract}
The weak boson fusion process for neutral Higgs boson production is 
investigated with particular attention to the accuracy with which the 
Higgs boson coupling to weak bosons can be determined at CERN Large 
Hadron Collider (LHC) energies in final states that contain a 
Higgs boson plus at least two jets.  Using fully differential perturbative 
matrix elements for the weak boson fusion signal process and for the 
QCD background processes, we generate events in which a Higgs boson is 
produced along with two jets that carry large transverse momentum.  The 
effectiveness of different prescriptions to enhance the signal to background 
ratio is studied, and the expected signal purities are calculated in each 
case.  We find that a simple cut on the rapidity of one final-state jet 
works well.  We determine that an accuracy of $\delta g/g \sim 10$\% on 
the effective coupling $g$ may be possible after 
$\sim 200$~fb$^{-1}$ of integrated luminosity is accumulated at the LHC.
\end{abstract}


\maketitle

\section{Introduction}
\label{sec:intro}

Following the discovery of the neutral Higgs boson $H$ at the CERN Large Hadron
Collider~(LHC), attention will focus on the measurement of its couplings
to gauge bosons and fermions.  A promising reaction from which
to extract some of these couplings, particularly the $HWW$ coupling, is 
the weak-boson fusion (WBF) 
process~\cite{Kinnunen:1999ak}--\cite{Cavalli:2002vs}, where the 
Higgs boson $H$ is produced via fusion of the weak bosons $W$ and $Z$: 
$WW, ZZ \to H$, and is accompanied in the final state by two
jets that carry large transverse momentum $p_T$.  To extract the 
couplings reliably, a good understanding is required of the 
production processes and the background processes that lead 
to the same final state.  Many strong interactions subprocesses also 
generate Higgs boson-plus-two-jet ($H+2$~jet) final states.  These 
background subprocesses can be computed with the techniques of perturbative 
quantum chromodynamics~(QCD).  They supply an {\em irreducible} background 
that may be reduced to some extent by judicious selections on the final 
state event topology.  

In the analysis presented here, we have in mind a situation in which  
the Higgs boson has been discovered and a sample of events exists 
containing a Higgs boson and two or more jets.  This set of events will 
contain backgrounds of two types: real $H+2$~jet events produced by QCD 
mechanisms other than WBF, and events which contain 
jets and particles that are present in typical Higgs boson decay modes, 
but without an explicit Higgs boson.  Within the full event sample, we discuss 
the simulation of the real WBF signal and the irreducible QCD $H+2$~jet 
background.  We do not address the second type of contamination, such as 
events from the QCD $Z+2$~jets process where the $Z$ decay imitates a Higgs 
boson decay.  Our concern is to estimate the expected signal purity, by which 
we mean the fraction of real Higgs boson events produced by weak boson fusion.

The WBF $H+2$~jet signal region is characterized by jets that carry large 
transverse momentum and large rapidity.  Because the jets carry large 
transverse momentum, it is necessary to use hard QCD matrix elements in order 
to represent the signal and the $H+2$~jet background reliably.  A parton shower 
approach, for example, would be unlikely to provide a correct estimate of the 
momentum distribution of the jets in the region of phase space of interest.    
Next-to-leading order (NLO) QCD corrections to the total WBF production 
cross section have been known for some time~\cite{Han:1992hr}, and  
the corresponding corrections were calculated recently in a 
fully differential way~\cite{Figy:2003nv}.  In this paper, we use an 
independent calculation to verify the results of Ref.~\cite{Figy:2003nv} 
and to examine in more detail the effects of the WBF selection cuts on 
the NLO QCD corrections.  We also use perturbative QCD expressions for 
the background $H+2$~jet matrix elements.  At present, the fully differential 
$H+2$~jet background distributions are known only at leading order. 
In addition to our NLO study of the signal process, we nevertheless provide 
two estimates of the NLO enhancement of the QCD $H+2$~jet background process, 
in order to better assess the viability of the WBF channel for measuring the 
coupling strength of the Higgs boson to vector bosons.  Our calculations are 
fully differential at the 
partonic level.  One limitation of the fact that we omit showering is that 
forward beam jets, which likely have low $p_T$, are ignored.  

Since the WBF channel is most interesting for a Higgs boson in the
mass range $m_H=115-200$~GeV, we perform calculations with the two 
extremal values of this range.  We compute differential cross sections 
in rapidity and transverse momentum at a $pp$ collider with 
$\sqrt{s}=14$~TeV. In Sec.~\ref{sec:MCimpl}, we discuss the production 
processes that contribute to the WBF signal and backgrounds, and we describe 
our method for evaluating them.  We generate momentum distributions using the 
general purpose Monte Carlo program MCFM~\cite{Campbell:2002tg}.  In our case, 
all jets carry a minimum value of relatively large transverse momentum whose 
values we specify.  

We present numerical values of the differential cross sections in
Sec.~\ref{sec:results}.  Various prescriptions are used in the literature to 
define the WBF sample, cuts that enhance the WBF fraction of the cross section 
by exploiting the special character of WBF events.  Our investigations lead us 
to propose a new, somewhat simpler definition in terms of a cut on the rapidity 
of one of the final state jets.  In this section, we also define quantitatively 
what we mean by WBF signal purity $P$.  We find that purities of 60\% to 70\% 
can be expected if a selection of $p_T \ge 40$~GeV is made on the tagging jets 
and somewhat lower values if the cut is dropped to $20$~GeV.  We derive an 
expression for the expected uncertainty on the effective 
Higgs-boson-to-weak-boson coupling strength $g$ in 
terms of $P$, the expected statistical accuracy of LHC experiments, and the 
uncertainties on the signal and the background processes.  We estimate that 
it should be possible to achieve an accuracy of $\delta g/g \sim 10$\% 
after $\sim 200$~fb$^{-1}$ of integrated luminosity is accumulated at the LHC.         
Somewhat smaller values of $\delta g/g$ are obtained in another 
recent investigation of anticipated uncertainties in the 
couplings~\cite{Duehrssen}, and we explain the source of the difference.  

In Sec.~\ref{sec:alternatives}, we compare the effects on both event 
rates and signal purity of our proposed method for defining WBF events 
with two other popular methods found in the literature: a selection on 
the difference in rapidities between two tagging jets in the final state, 
and a selection on the invariant mass of a pair of tagging jets.  The 
alternative prescriptions yield some increase in signal purity with respect 
to our definition, but the gain is sensitive to the cut in transverse 
momentum used to specify the trigger jets, and it is accompanied by loss 
of event rate.  For values of the jet cut $p_T > 40$~GeV, our prescription appears 
to work about as well as the other methods.  Relatively high luminosity will 
be needed for a precise determination of the uncertainty $\delta g/g$.  Our 
simpler definition of the WBF sample in terms of a selection on the rapidity 
of only one jet offers advantages in a high luminosity environment where a 
large value of the transverse momentum cut is appropriate and multiple events 
per crossing may be an issue.

We provide two methods for estimating the size of next-to-leading order 
corrections to the $H + 2$~jet background in Sec.~\ref{sec:estimating}.  One 
of these relies on similarity with the $Z + 2$~jet process for which fully 
differential NLO results are known.  The other method is an extrapolation 
from the known next-to-next-to-leading order (NNLO) results for the fully 
inclusive Higgs boson production 
process.  The substantially different estimates for the NLO enhancement 
provided by these two methods show the level of uncertainty of the LO 
background calculation.  
A fully differential NLO calculation of the $H + 2$~jet 
background applicable in the region of interest for WBF investigations is 
needed in order to improve our computations of signal purity and of the 
expected uncertainty in $\delta g/g$.  A summary of our conclusions may be 
found in Sec.~\ref{sec:conclusions}.   

\section{Production Cross Sections}
\label{sec:MCimpl}

\begin{figure}[h]
\begin{center}
\epsfig{file=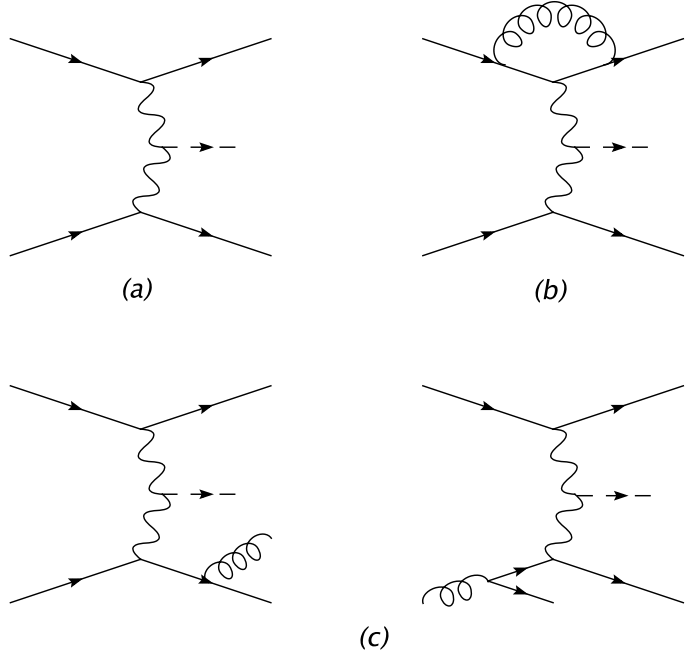,angle=0,width=9cm} \\
\end{center}
\caption{Representative diagrams for the production of a Higgs boson
via weak boson fusion: (a) at lowest order; (b),(c) at NLO. Further
diagrams can be obtained by crossing incoming and outgoing lines 
in all cases. All of the virtual
corrections are of the vertex correction form, as shown in (b). There
are two types of real corrections depicted in (c). The first set
corresponds to the emission of a gluon in all possible positions on the
quark lines (left-hand diagram) and the second set corresponds to the
crossing where a gluon is present in the initial state (right-hand
side).}
\label{fig:wbfdiags}
\end{figure}

Examples of the WBF diagrams that must be calculated are shown in 
Fig.~\ref{fig:wbfdiags}.
The basic leading order process is shown in (a), where the exchanged
bosons may be either $W$'s or $Z$'s, and one or both quark lines may
be reversed, yielding $qq$, $q{\bar q}$ and ${\bar q}{\bar q}$ initial
states. The virtual NLO corrections are obtained by adding a gluon
loop to either $qqV$ vertex, as illustrated in (b). The remaining real
NLO corrections are shown in (c), where either an additional gluon
is radiated in the final state or a gluon from the proton splits into
a $q{\bar q}$ pair.  Calculation of the necessary loop diagrams is 
straightforward, providing a couple of simplifying assumptions are made. 
First, we ignore contributions of the form 
$q {\bar q}' \to V^{\star} \to VH$,
where $V=W,Z$. Second, we neglect any interference effects from identical 
flavor quarks in the final state.  We checked that both of these 
approximations have little effect on the calculated cross sections at 
leading order, particularly in the region of phase space that we consider.

The NLO calculation is embedded in the general purpose Monte Carlo
program MCFM~\cite{Campbell:2002tg}, which uses the dipole subtraction
method~\cite{Catani:1996vz}. We use the default set of parameters in
this program, in which $\alpha=1/128.89$, $M_W=80.419$~GeV, $M_Z=91.188$~GeV, 
and $\sin^2\theta_w=0.2285$.  For the parton distribution
functions, we use  CTEQ6L1 for lowest order and CTEQ6M at
NLO~\cite{Pumplin:2002vw}.  In these sets of parton densities, 
$\alpha^{\rm LO}_s(M_Z)=0.130$ and $\alpha^{\rm NLO}_s(M_Z)=0.118$.
In this paper, we choose the reference value $\mu=m_H$ for the 
renormalization and factorization scales.  In Sec.~\ref{sec:mudep}  
we investigate the uncertainty of the signal and of the background 
associated with variation of the scale over the interval 
$2m_H > \mu > m_H/2$.

\subsection{Generic cuts}
The hallmark of WBF events is a Higgs boson accompanied by two 
``tagging'' jets having large $p_T$ and large rapidity.  In real events and 
in computations at 
NLO, there are generally more than two jets, and the goal is to pick 
out a clean signal.  To simplify our study and to demonstrate the robust 
character of the WBF process, it is desirable to make as few selections 
(cuts) as possible on the events.  We begin with a basic set of cuts, 
exactly as in Ref.~\cite{Figy:2003nv}.  Jets obtained from the Monte Carlo 
runs are clustered according to the $k_T$ algorithm with 
$p_T^{\rm jet} > 20$~GeV, jet pseudo-rapidity $|\eta^{\rm jet}| < 4.5$,  
and jet separation 
$\Delta R_{jj}=\sqrt{ \Delta\eta_{jj}^2 + \Delta\phi_{jj}^2} > 0.8$, 
where $\Delta\phi_{jj}$ is the difference in the azimuthal angles of the 
two jets in the transverse plane.  
The two jets with the highest $p_T$ are chosen as the tagging jets and
ordered according to their pseudo-rapidities, $\eta_{j_1} < \eta_{j_2}$.
In order to approximate the acceptance for the Higgs boson decay products
and to complete the specification of our minimal set of cuts, 
we imagine the decay of a Higgs boson to two charged particles, denoted 
as ``leptons''.  We require that these leptons satisfy the 
cuts:
$$
p_T^{\rm lept} > 20~{\rm GeV}, |\eta^{\rm lept}| < 2.5,
\Delta R_{j\ell} > 0.6, \eta_{j_1} < \eta_{\rm lept} < \eta_{j_2}.
$$
The Higgs boson decay products are therefore located in pseudo-rapidity 
between the two high-$p_T$ jets.  Although we enforce these cuts on 
potential Higgs boson decay products, the cross sections that we present 
do not include any branching ratio for this decay.  We include branching 
ratios and efficiencies when we discuss the determination of the coupling 
strength in Sec.~\ref{sec:results}.

\begin{figure}[h]
\begin{center}
\epsfig{file=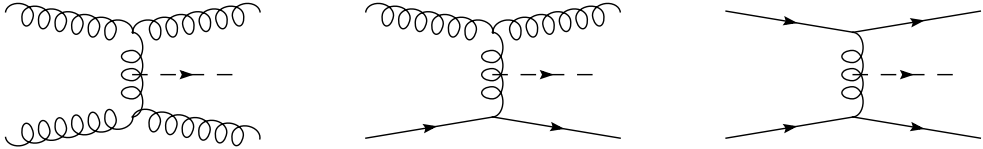,angle=0,width=13cm} 
\end{center}
\caption{Representative diagrams for the production of a Higgs boson
and two jets at lowest order, calculated in the heavy top-quark limit
of the $Hgg$ effective coupling.}
\label{fig:h2jdiags}
\end{figure}

Throughout this paper we refer to the QCD production of a Higgs
boson in association with jets as the ``background'' to our WBF
signal events. This cross section is implemented in MCFM at leading
order based on the matrix elements of Ref.~\cite{Kauffman:1996ix}.  
A selection of the contributing diagrams is shown in
Fig.~\ref{fig:h2jdiags}.  There are contributions from $qq$, $qg$, 
and $gg$ initial state subprocesses, but the Higgs boson is always 
produced from an effective $gg \to H$ vertex.  
The effective coupling of the Higgs boson to two gluons is included in
the limit of heavy top-quark mass $m_t$, with a coupling strength 
$\alpha_s/(3\pi v)$, where $v$ is the Higgs boson vacuum expectation value, 
and $\alpha_s$ is evaluated at the scale $m_H$ with the $1$-loop 
expression for the evolution ($0.125$ for $m_H=115$~GeV and $0.116$ for
$m_H=200$~GeV).  The effective coupling approximation should be valid since 
we limit ourselves to Higgs boson masses $m_H < 2m_t$ and Higgs boson 
transverse momenta $p_T^H < m_t$ ~\cite{DelDuca:2001fn}.

In Sec.~\ref{sec:estimating} we estimate the NLO corrections to the  
lowest order result for the $H+2$~jets background process by comparison 
with the similar $Z+2$~jets process, calculated at NLO~in 
Ref.~\cite{Campbell:2002tg,Campbell:2003hd}, and by extrapolation from 
NNLO calculations of the fully inclusive process 
$p p \rightarrow H X$~\cite{Catani:2001ic}--\cite{Berger:2003pd}, 
and NLO calculations of 
$p p \rightarrow H + 1~{\rm jet} + X$~\cite{Ravindran:2002dc}.

\section{Results}
\label{sec:results}

The cuts mentioned in the previous section are a generic set of 
cuts.  They do not exploit the kinematic structure of WBF events, 
where the jets tend to be produced very forward in pseudo-rapidity. 
In this subsection, we present first the cross sections for the 
signal and background processes after application of the generic 
cuts.  Without further cuts, the WBF events would be lost in the 
QCD continuum background. We then apply one additional constraint 
which defines our WBF sample, and we show results for kinematic 
distributions, event rates, and signal purities.

\subsection{Basic cuts}
\label{sec:WBFsig}
We examine the effects of the generic cuts in terms of
their effects on the WBF signal and the $H+2$~jet background. These
cross sections -- without any further cuts -- are shown in
Table~\ref{tab:rates_ptdep} as a function of the minimum jet $p_T$.  
The WBF signal process is calculated at NLO and, at this point, the 
$H+2$~jet background at LO.
\begin{table} 
\begin{center} 
\begin{tabular}{|l|c|c|c|} \hline 
$p_T$ cut [GeV]& $20$ & $40$ & $80$ \\ \hline 
Signal ($m_H=115$)   & 1866 & 1081 & 239 \\ \hline 
Bkg                  & 2173 & 743 & 200 \\ \hline \hline 
Signal ($m_H=200$)   & 1189 &  709 &166 \\ \hline 
Bkg                  &  958 &  340 &  96 \\ \hline
\end{tabular} 
\caption{Cross sections in fb for the WBF signal(calculated at NLO) and
$H+2$~jet background(LO), as a function of the minimum jet $p_T$.
Only the minimal set of cuts of Sec.~\ref{sec:MCimpl} is applied.}
\label{tab:rates_ptdep}
\end{center}
\end{table}
We remark that the effects of the NLO corrections on the WBF process
are rather small in this region, corresponding to $K$-factors between
$0.95$ and $1.1$, depending on the Higgs boson mass and $p_T$ cut. 

The values in the table show that (without consideration yet of NLO effects 
in the background) the rates for the signal and background are comparable 
for $p_T^{\rm min}=20$~GeV, and the signal-to-background ratio improves as
the $p_T$ cut is increased.  The WBF signal lies above the $H+2$~jet cross 
section if $p_T^{\rm min} \ge 40$~GeV.

\subsection{WBF cuts}
\label{sec:ourcuts}
\begin{figure}[h]
\begin{center} 
\epsfig{file=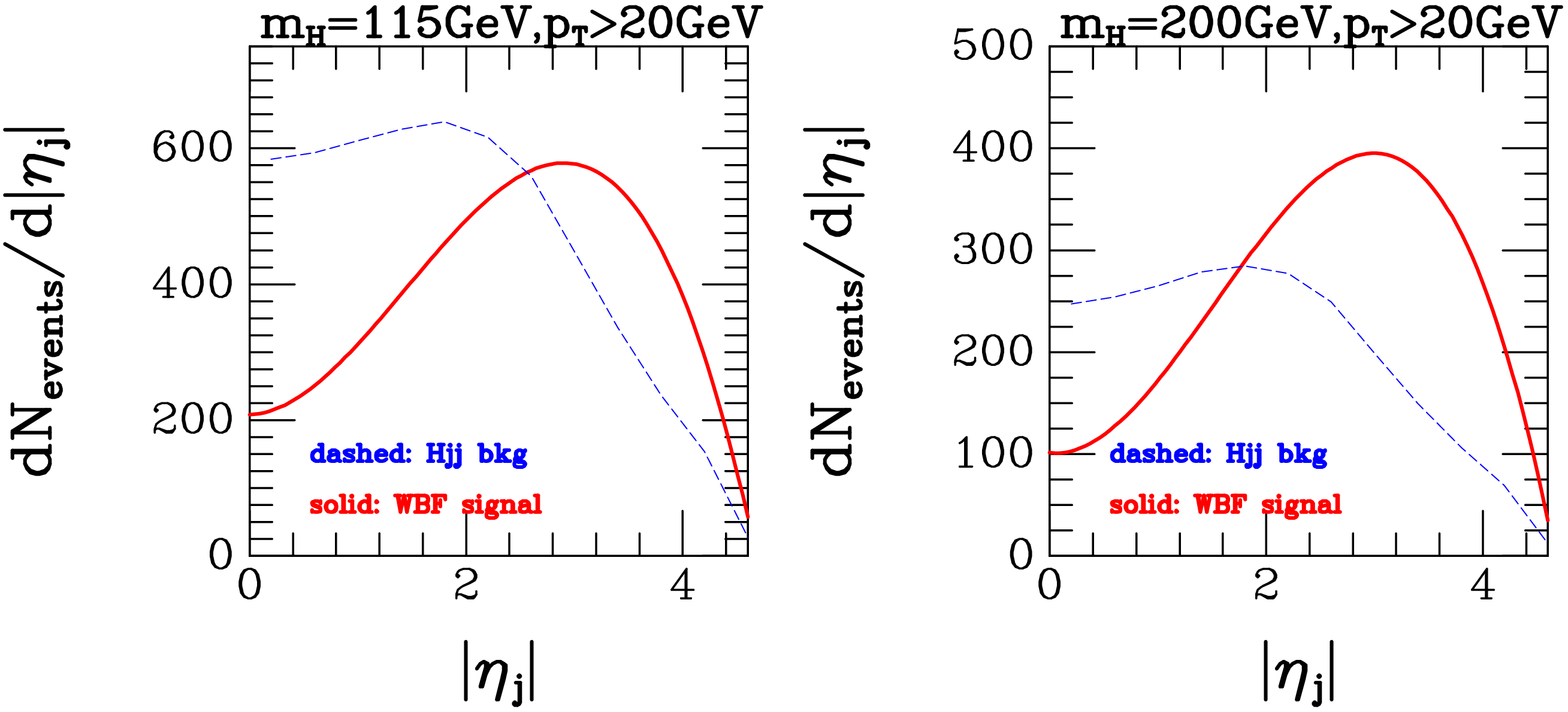,angle=270,width=13cm} \\ \vspace*{0.3cm}
\epsfig{file=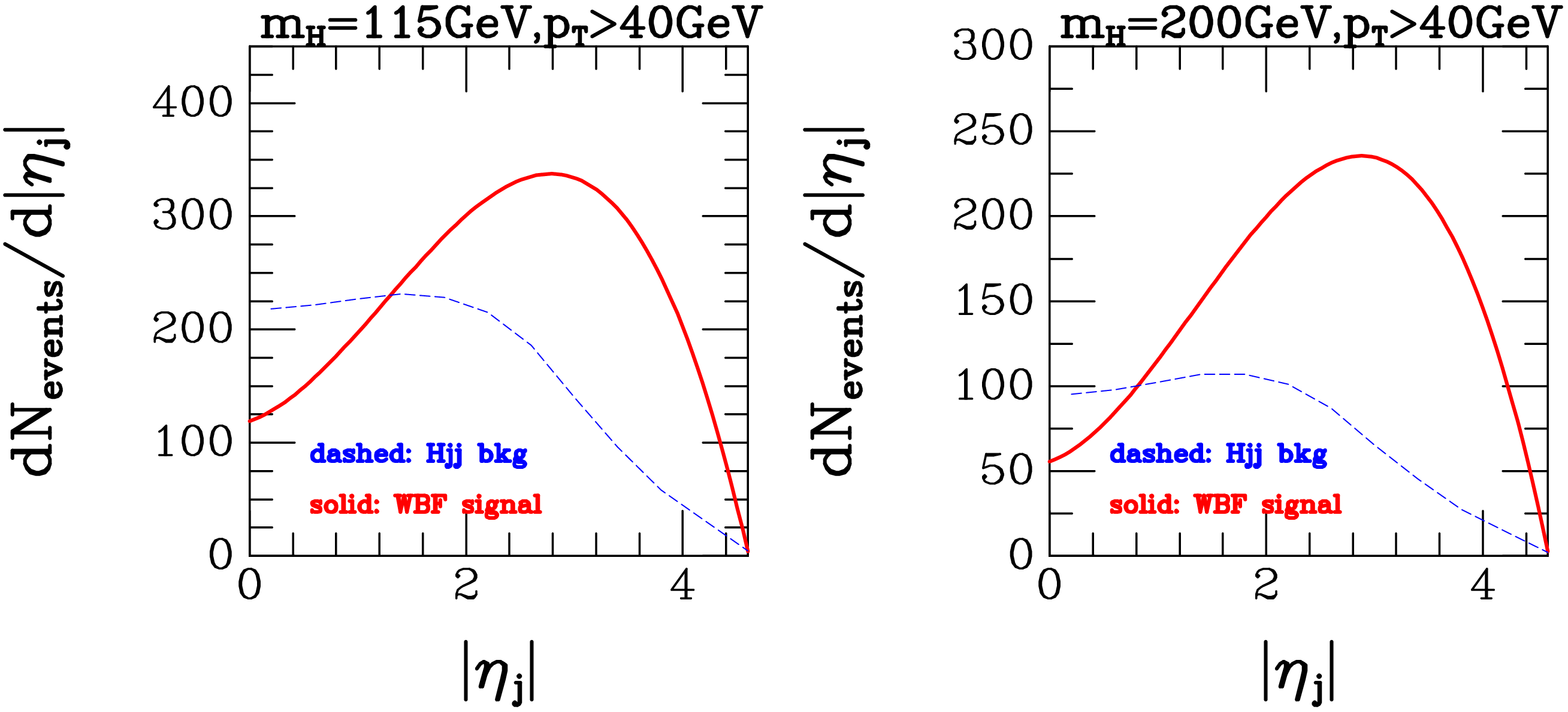,angle=270,width=13cm} \\ \vspace*{0.3cm}
\epsfig{file=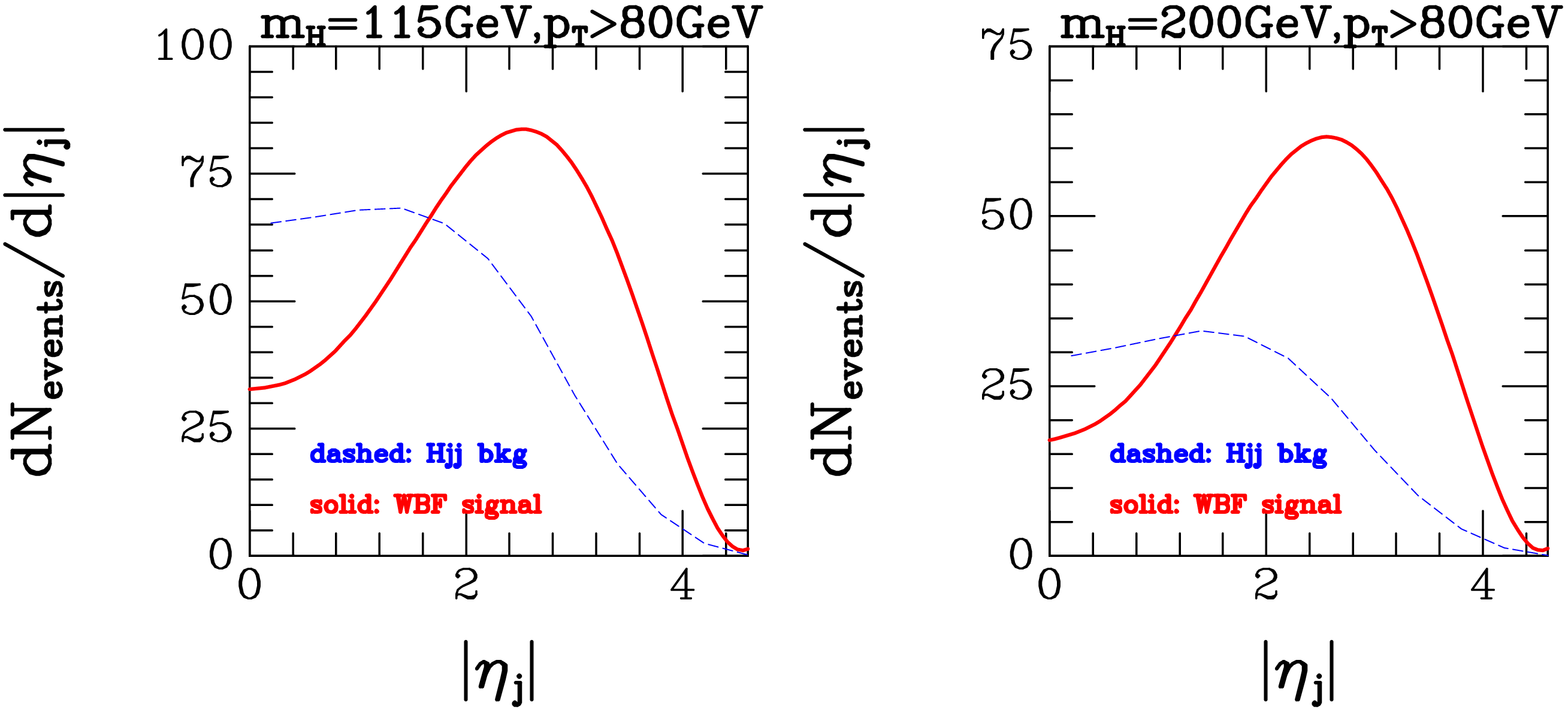,angle=270,width=13cm} \\ \vspace*{0.3cm} 
\end{center}
\caption{Dependence of the tagging jet pseudo-rapidities on the minimum 
jet $p_T$ used, for the two cases $m_H=115$~GeV (left) and $m_H=200$~GeV
(right). Each of the two tagging jets in the event is entered in these
plots, with weight one-half, and the rates assume an integrated
luminosity of $1~{\rm fb}^{-1}$. The signal (solid) is calculated at
NLO and the background (dashed) at LO.  We show results for three 
different selections of the minimum jet transverse momentum, $p_T > 
20$, $40$, and $80$~GeV.}
\label{fig:absrap_pt}
\end{figure}

In an attempt to exploit the WBF event structure, a popular cut invokes a 
separation in pseudo-rapidity between the two tagging jets, for instance
$|\eta_{j_1}-\eta_{j_2}|>4$. In this paper, we define a slightly
different and simpler cut, motivated by our examination of the 
distributions of the absolute jet pseudo-rapidities shown in 
Fig.~\ref{fig:absrap_pt}.
In these figures, each tagging jet enters with weight one-half and
cross sections have been converted to event rates with an integrated
luminosity of $1~{\rm fb}^{-1}$. The area under each curve is equal
to the total number of events in that channel. 

The plots in Fig.~\ref{fig:absrap_pt} show that the shape of the distribution 
depends little on either the Higgs boson mass or the jet $p_T$, but -- 
as expected -- is very different in the signal process, compared to the 
$H+2$~jet background.  In each case, the WBF events peak at values of 
$|\eta| \approx 3$, although there is a slight movement to lower values 
of $|\eta|$ as the $p_T$ cut is increased.  The width of the peak also 
tends to decrease, but the full width at half-maximum is fairly constant
at approximately $3$ units of rapidity. In contrast, the rate of
background events falls off fairly sharply beyond $|\eta| \approx 2$.  

Motivated by the comparison of rapidity spectra in Fig.~\ref{fig:absrap_pt}, 
and erring on the side of simplicity, we choose a uniform cut that ensures  
at least one jet lies within the peak. Namely,
\begin{equation}
\eta_{\rm peak}-\eta_{\rm width}/2 < |\eta_{j}| <
 \eta_{\rm peak}+\eta_{\rm width}/2,
\label{etapeak}
\end{equation}
for $j=j_1$ or $j=j_2$, where $\eta_{\rm peak}$=3 and
$\eta_{\rm width}$=2.8. 

\begin{figure}[h]
\begin{center}
\epsfig{file=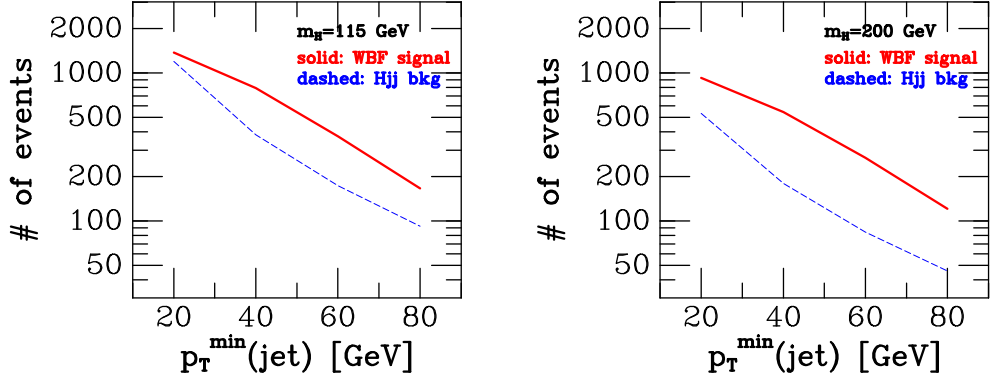,angle=270,width=13cm} \\
\end{center}
\caption{Numbers of events for the WBF signal and the QCD background as a
function of the minimum jet $p_T$, for an integrated luminosity of
$1~{\rm fb}^{-1}$. No branching ratios for the Higgs boson decay have been
applied. The pseudo-rapidity restriction on one of the jets has been enforced,
as in Table~\ref{tab:rates_rapsep}. The solid line is the NLO signal
and the dashed is the LO background.}
\label{fig:rates_rapsep}
\end{figure}

{\em Equation~(\ref{etapeak}), along with the 
generic cuts specified above, constitutes our definition of weak 
boson fusion cuts.}  The effects of the pseudo-rapidity restriction on 
the jets are shown in Table~\ref{tab:rates_rapsep}, where we have
assumed $1~{\rm fb}^{-1}$ of integrated luminosity.  The rates in this 
table should be contrasted with those in Table~\ref{tab:rates_ptdep}.  
In Table~\ref{tab:rates_rapsep}, we include values for the signal purity, 
defined as $P = S/(S+B)$, where $S$ stands for the number of signal 
events and $B$ for the number of background events.
\begin{table}
\begin{center}
\begin{tabular}{|l|c|c|c|} \hline
$p_T$ cut [GeV]      & $20$ & $40$ & $80$ \\ \hline
Signal ($m_H=115$)   & 1374 &  789 &  166 \\ \hline
Bkg                  & 1196 &  382 &   92 \\ \hline
Purity               & 0.53 & 0.67 & 0.64 \\ \hline \hline
Signal ($m_H=200$)   &  928 &  545 &  121 \\ \hline
Bkg                  &  534 &  179 &   46 \\ \hline
Purity               & 0.63 & 0.75 & 0.72 \\ \hline
\end{tabular}
\caption{
Event rates for the $Hjj$ WBF signal(NLO) and 
$Hjj$ background(LO), including our WBF requirement that at least 
one jet carry large $|\eta|$, as defined by Eq.~(\ref{etapeak}).  
We assume $1~{\rm fb}^{-1}$ of integrated luminosity.  Purity is 
defined as $P = S/(S+B)$, where $S$ stands for the number of signal 
events and $B$ for the number of background events.}
\label{tab:rates_rapsep}
\end{center} 
\end{table}
The number of events as a function of the minimum jet $p_T$ is
also plotted in Fig.~\ref{fig:rates_rapsep}.
Comparing the tables, one can see that the signal rate is diminished 
only slightly, by about $20$--$30\%$. On the other
hand, the background is shrunk considerably, by about a factor of two.
A $p_T$ cut of $20$~GeV is barely sufficient to distinguish the WBF signal 
above the QCD LO $Hjj$ background for $m_H=115$~GeV.  However, the signal 
$S$ to background $B$ ratio improves to about 2 for 
$p^{\rm cut}_T \ge 40$~GeV.  At $m_H=200$~GeV, the situation is better, 
with $S/B$ of about $1.7$ when the $p_T$ cut is $20$~GeV, and rising 
to $\sim 3$ for $p^{\rm cut}_T \ge 40$~GeV. A $p_T$ cut of $40$~GeV yields 
a prominent effect across the range of interesting masses, 
$m_H=115$--$200$~GeV.  

\begin{figure}[h]
\begin{center} 
\epsfig{file=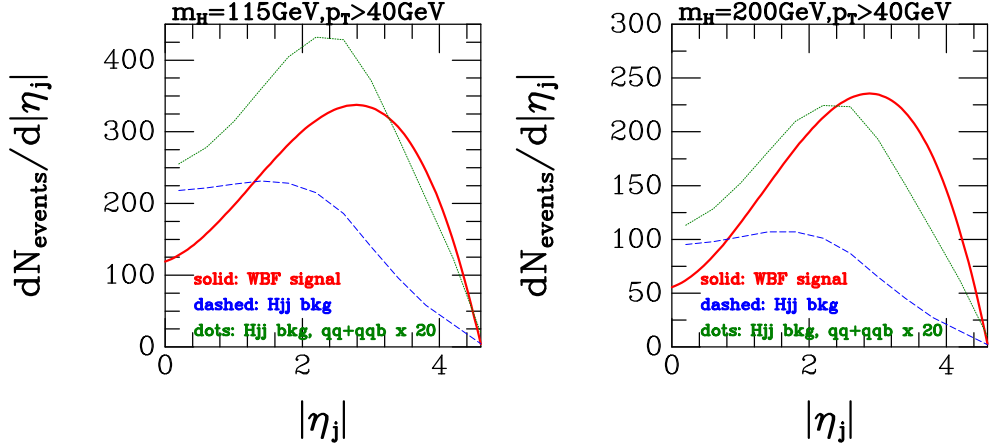,angle=270,width=13cm}
\end{center}
\caption{Dependence of the tagging jet pseudo-rapidities for jet 
$p_T > 40$~GeV, for the two cases $m_H=115$~GeV (left) and $m_H=200$~GeV
(right). For the background, we show the 
full result with all contributions included and, for comparison of 
shapes, the background obtained if only the $qq$, $q \bar{q}$, and 
$\bar{q} \bar{q}$ 
initial state contributions are used.  The magnitude of the separate 
component is multiplied by $20$.}
\label{fig:absrap_pt_qqonly}
\end{figure}

It is instructive to examine the origin of the different rapidity 
spectra for the signal and the background.  Since $H+2$~jet events are 
generated in both cases with identical cuts on the transverse momenta 
of the jets, the different rapidity spectra must originate from dynamics.  
Comparing the LO production diagrams in Figs.~\ref{fig:wbfdiags} 
and~\ref{fig:h2jdiags}, we note that $gg$ and $qg$ initial states 
contribute to the QCD background but not to the WBF signal.  The 
gluon parton density is notably softer than the quark parton density, 
suggesting a plausible reason for the differences in the rapidity 
spectra of the final state jets in the two cases.  This reasoning is 
supported by the results shown in Fig.~\ref{fig:absrap_pt_qqonly}.  
The shape of the background rapidity spectrum from the $qq$, 
$q \bar{q}$, and $\bar{q} \bar{q}$ contributions is very similar to 
that of the signal, albeit with a slight shift of the peak to smaller 
$|\eta|$.  The very different rapidity spectra of the signal and 
the background evident in Fig.~\ref{fig:absrap_pt} results therefore 
primarily from the $gg$ and $qg$ initial state contributions.  The 
results shown in Fig.~\ref{fig:absrap_pt_qqonly} imply that there is 
a basic upper limit to the purity one can achieve for the WBF event 
sample, regardless of which prescription one adopts to define the WBF 
sample.  The $qq$, $q \bar{q}$, plus $\bar{q} \bar{q}$ component of 
the QCD background process generates a final state event topology 
essentially identical to the WBF signal process.  Values of the 
purity are listed Table~\ref{tab:rates_rapsepqq}; there is not much 
variation with $m_H$ or the value of the cut in $p_T$.  Our results 
suggest that purity is bounded from above by at most $P < 0.95$ 
at LHC energies.  

\begin{table}
\begin{center}
\begin{tabular}{|l|c|c|c|} \hline
$p_T$ cut [GeV]      & $20$ & $40$ & $80$ \\ \hline
Signal ($m_H=115$)   & 1374 &  789 &  166 \\ \hline
Bkg ($qq$,$q\bar{q}$,$\bar{q}\bar{q}$)   & 98 &  45 &  15 \\ \hline
Purity ($qq$,$q\bar{q}$,$\bar{q}\bar{q}$) & 0.93 & 0.95 & 0.92 \\ \hline \hline
Signal ($m_H=200$)   &  928 &  545 &  121 \\ \hline
Bkg                  &   47 &   23 &    8 \\ \hline
Purity               & 0.95 & 0.96 & 0.94 \\ \hline
\end{tabular}
\caption{
Event rates for the $Hjj$ WBF signal(NLO) and for the part of the 
$Hjj$ background(LO) that arises from the $qq$, $q\bar{q}$, and 
$\bar{q}\bar{q}$ initial-state terms.}
\label{tab:rates_rapsepqq}
\end{center} 
\end{table}

\subsection{Scale dependence study}
\label{sec:mudep}

\begin{figure}[h]
\begin{center} 
\epsfig{file=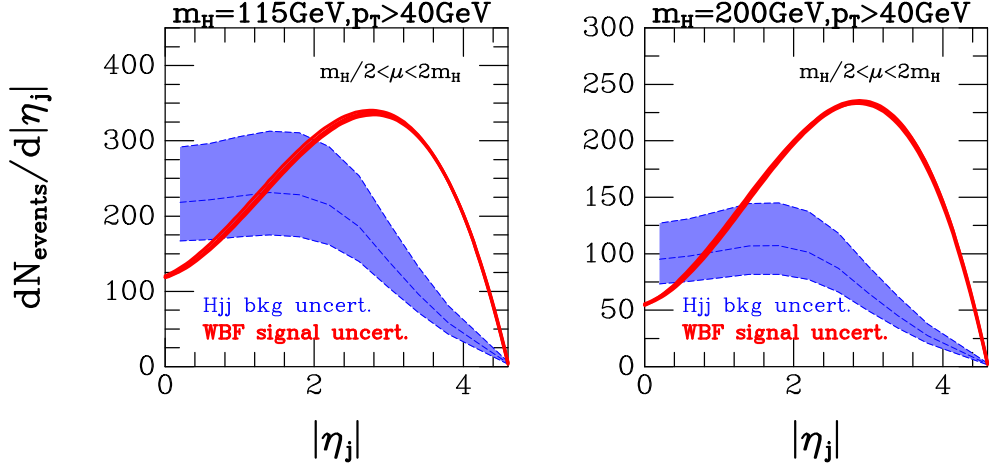,angle=270,width=13cm} \\
\end{center}
\caption{Tagging jet pseudo-rapidity distributions in $Hjj$ events, calculated 
with a range of values of the renormalization and factorization scale, $\mu$.  
The signal (solid) is calculated at NLO and the background (dashed) at LO.}
\label{absrap_scale}
\end{figure}

To examine further the effects of NLO corrections, we consider variation of 
the renormalization and factorization scale.  A range of values 
$m_H/2 < \mu < 2m_H$ is used conventionally to estimate the theoretical 
uncertainty at a given order in perturbation theory.  As a representative case, 
we show results for a minimum jet $p_T$ of $40$~GeV and both Higgs boson 
masses.  
In Fig.~\ref{absrap_scale}, we show the tagging jet pseudo-rapidity
distributions for the signal and background for a range of values of  
the common renormalization and factorization scale $\mu$.  The signal 
process shows very little variation with $\mu$, a shift of less than 
$\pm 2$\% when $|\eta_j| \sim 2$ in the WBF 
signal region.  In contrast, 
the $H+2$~jet background at LO is enhanced by approximately $70\%$ when the 
scale choice $\mu = m_H/2$ is made, and reduced by $40$\% for $\mu = 2m_H$.  A 
fully differential NLO calculation of the $H+2$~jet background process is required 
to reduce the large uncertainty associated with $\mu$ variation apparent in 
Fig.~\ref{absrap_scale}.

\subsection{Uncertainty on the Couplings}
\label{sec:uncert}

The signal and background events both include a real Higgs boson along with 
two jets.  We may define a signal ``purity'' as the ratio $S/(S+B)$, where 
$S$ denotes the number of signal events and $B$ the number of background 
events.   The purity as defined here does not improve with greater luminosity 
nor does it depend on the Higgs boson decay mode considered.  
Of interest to us is the effect of signal purity on the accuracy of the 
determination from data of the Higgs boson couplings $g_{WW}$ and $g_{ZZ}$
to the $WW$ and $ZZ$ channels.  The WBF cross section is proportional to a 
combination of $g_{WW}^2$ and $g_{ZZ}^2$, and their relative contribution 
changes somewhat with the value of the cut on $p_T$.  In this paper, we discuss 
only an effective coupling strength $g$.  We remark also that in our discussion 
of the expected accuracy on $g$, we limit ourselves to uncertainties at the level 
of production of the Higgs boson.  We set aside uncertainties associated with the 
fact that the Higgs boson is observed only in specific final states and that 
all the final states cannot be observed above backgrounds. 

To derive the uncertainty $\delta g/g$ on the coupling, we begin with the 
observed number of events $N = S + B$.  We define 
the ratio $r = g^2_{\rm observed}/g^2_{\rm predicted}$.  Then, under the 
assumption that any deviation in the expected total number of events arises 
from the effective coupling, we obtain $r = (N-B)/S$.  Taking the total derivative, 
we obtain an expression for the uncertainty in $r$.  
\begin{equation}
\delta r/r = \sqrt{[ (\delta S/S)^2 + ((\delta N)^2 + (\delta B)^2)/(N-B)^2 ]} ,
\label{uncert1}
\end{equation}
and, correspondingly,
\begin{equation}
\delta g /g = 1/2 \sqrt {[ (\delta S/S)^2 + ((\delta N)^2 + (\delta B)^2)/(N-B)^2 ]} .
\label{uncert2}
\end{equation}
With purity $P = S/(S+B)$, we derive 
\begin{equation}
\delta g /g = 1/2 \sqrt {[ (\delta S/S)^2 + (1/P)^2(\delta N/N)^2 + ((1-P)/P)^2(\delta B/B)^2 ]} .
\label{uncert}
\end{equation}

In the absence of any uncertainty in knowledge of the signal and background, 
Eq.~(\ref{uncert}) demonstrates the obvious fact that the best one can achieve 
is $\delta g/g = 0.5 \, \delta N/N$ for $P  = 1$.  
The factor $(1/P)$ that multiplies $\delta N/N$ in Eq.~(\ref{uncert}) shows  
that reduction in the purity effectively reduces the statistical power of the data. 
Similarly, the factor $(1-P)/P$ that multiplies $\delta B/B$  shows that greater purity 
diminishes the role of uncertainty in our knowledge of the background.  Given that 
purity decreases as the background increases, we see that the size of the background 
in the WBF region is the problem to contend with; the uncertainty on the background is of 
less importance.  To represent the background reliably in a region of phase space in 
which tagging jets carry large transverse momentum, it is clearly important to use 
partonic hard matrix elements that can simulate this jet activity.   
\begin{figure}[h]
\begin{center}
\epsfig{file=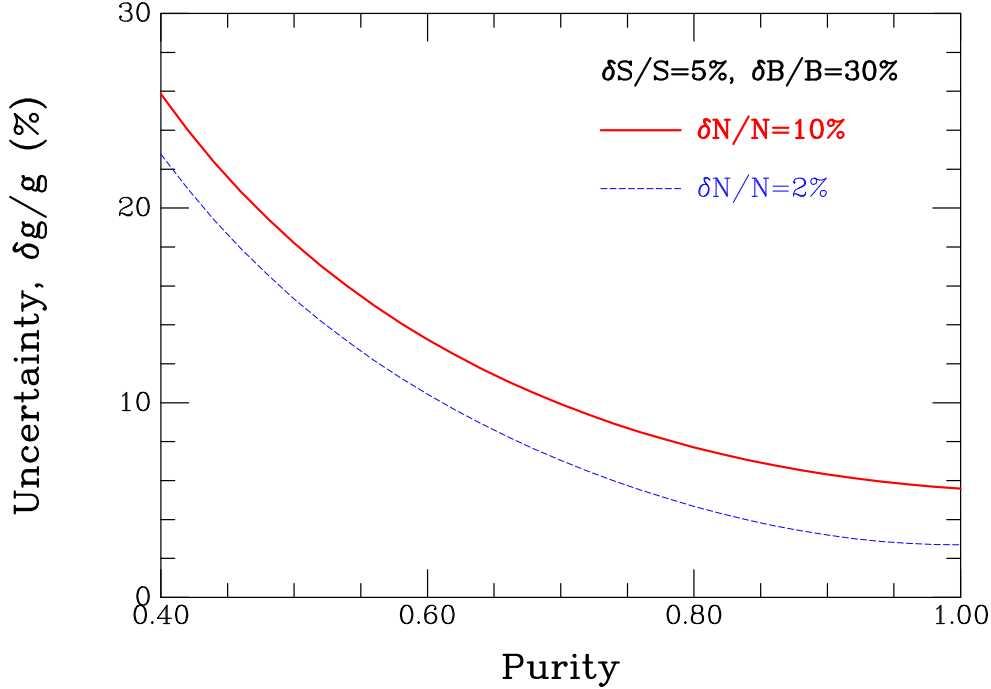,angle=270,width=13cm} \\
\end{center}
\caption{The predicted uncertainty $\delta g/g$ in the coupling of the Higgs 
boson to a pair of $W$ bosons is shown as a function of signal purity 
$P = S/(S+B)$ for expected statistical accuracies $\delta N/N$ of 10\% and 2\% 
The uncertainties in knowledge of the signal $S$ and background $B$ are
assumed to be 5\% and 30\% respectively.}
\label{fig:uncert}
\end{figure}

To obtain numerical values for the overall uncertainty in $g$, we must specify the 
uncertainties expected in our knowledge of the signal $S$ and the background $B$, along 
with the statistical uncertainty $\delta N/N$.  We address each of these contributions.  

The NLO QCD effects on the WBF signal are modest as are uncertainties associated with 
parton densities and variation of the renormalization/factorization scale. We may take 
$\delta S/S = 5$\%, based on the calculated NLO $\mu$ dependence of $\pm 2$\% and 
PDF uncertainty of $\sim 3$\%, both obtained in the WBF region of phase-space.  
Next-to-leading order QCD contributions to the background are discussed in 
Sec.~\ref{sec:estimating}.  The size of these contributions, scale dependence, and 
parton density variation could make the LO background estimate uncertain at the 60\% 
level.  We adopt a perhaps optimistic value of $\delta B/B = 30$\%.  This choice 
presupposes that the $20$\% $\mu$ variation and $5$\% PDF uncertainty of the fully 
inclusive NLO cross section for Higgs boson production may also apply to the NLO 
calculation of $H+2$~jet production in the WBF region of phase-space, once this 
calculation is done.     

Based on a study of conventional backgrounds~\cite{Asai:2004ws}, 
a minimum of roughly 10~fb$^{-1}$ of integrated luminosity is needed to discover the 
Higgs boson in the WBF process.  This figure would be achieved after one year of LHC 
operation at a luminosity of $10^{33}{\rm cm}^{-2}{\rm s}^{-1}$.  
Using the numbers in our Table~\ref{tab:rates_rapsep}, we expect a WBF sample ($S+B$) of 
$\sim 12000$ events for $m_H = 115$~GeV and $p_{Tcut} = 40$~GeV, and 
$\sim 7000$ events for $m_H = 200$~GeV and $p_{Tcut} = 40$~GeV.  
To translate these event rates into statements about statistical significance, we must 
specify a Higgs boson decay mode and approximate tagging efficiencies for the decay 
products.  For $m_H=115$~GeV, we choose the decay $H \to \tau^+ \tau^-$, with one 
$\tau$ decaying leptonically and the other hadronically~\cite{Rainwater:1998kj}.
These choices yield a branching ratio of 
$$Br(H \to \tau\tau) \times Br(\tau \to {\rm leptons})
\times Br(\tau \to {\rm hadrons})
 = 0.073 \times 0.7 \times 0.65 = 0.033.
$$
For the efficiency for tagging hadronic $\tau$ decays, we take the figure 
$0.26$ from Ref.~\cite{Cavalli:note} as an optimistic upper bound.{\footnote 
{We recall that the generic acceptance cuts defined in Sec.~\ref{sec:MCimpl} for 
the Higgs boson decay products are included in our event rates.}} The  
true value for this efficiency will be known only after analysis of data from 
LHC experiments. While our choice of tagging efficiency may seem large, it is 
relatively easy to scale our final results if a different value is preferred.    
The combination of branching fraction and efficiency results in a reduction 
in the number events by a factor $\epsilon = 0.033 \times 0.26 \approx 0.01$.  
For $m_H=200$~GeV, the decay $H \to W^+ W^-$ is prominent, and we select the case 
in which both $W$'s decay leptonically~\cite{Kauer:2000hi}.{\footnote {We 
acknowledge that $H \to W^+ W^-$ with leptonic decay of both $W$'s is not a 
perfect match to our earlier specification of two body decay of the Higgs boson to 
``leptons''.  Because the $W$ is fairly massive, the rapidity distribution of 
the decay leptons may extend beyond $|\eta| < 2.5$, and a further acceptance 
correction may have to be applied.  Such a study is best addressed by an 
experimental simulation of the entire decay chain.}}  We obtain 
$$
\epsilon=Br(H \to WW) \times Br(W \to {\rm leptons})^2
 = 0.74 \times 0.22^2 = 0.036.
$$
Using these numbers, we compute expected statistical uncertainties of  
$\delta N/N \sim 10$\% and $\sim 6$\% at $m_H = 115$ and $200$~GeV, respectively.   
With statistical accuracy $\delta N/N$ of 10\%, $\delta S/S = 5$\% and 
$\delta B/B = 30$\%, we obtain $\delta g/g \simeq 10$\% for purity $P = 0.7$ 
when $m_H = 115$~GeV, and $\delta g/g \simeq 8$\% when $m_H = 200$~GeV.  

After 5 years of LHC operation, we can anticipate an integrated luminosity of 
200~fb$^{-1}$ will have been accumulated.  This increase allows us to reduce our 
estimates of $\delta N/N$ to 
$\sim 2$\% and $\sim 1.5$\% at $m_H = 115$ and $200$~GeV, respectively, and 
$\delta g/g \sim 7$\% for $P =0.7$.  

We remark that the uncertainties in the signal $S$ and background $B$ dominate the 
uncertainty in $g$.  If $P = 0.7$ and $\delta N/N = 2$\%, then the uncertainties 
$\delta S/S$ and $\delta B/B$ would have to be reduced to 3\% and 6\%, respectively, 
before the statistical uncertainty would control the answer.  Even if $P = 1$, 
$\delta g/g$ is controlled by $\delta S/S$ until $\delta S/S \le \delta N/N$.  

In Fig.~\ref{fig:uncert}, we show numerical predictions for the uncertainty as a function 
of purity, for two choices of the statistical uncertainty.  Signal purities of $0.65$ or 
greater permit determinations of $\delta g/g$ of $10$\% or better after 200~fb$^{-1}$ have 
been accumulated.  As shown in Table~\ref{tab:rates_rapsep}, $P > 0.65$ is obtained 
for $p_{Tcut} > 40$~GeV at $m_H = 115$~GeV, and $p_{Tcut} > 20$~GeV at $m_H = 200$~GeV.  
The curves indicate to us that there is not much to gain from purities greater than $70$\%.

\begin{figure}[h]
\begin{center}
\epsfig{file=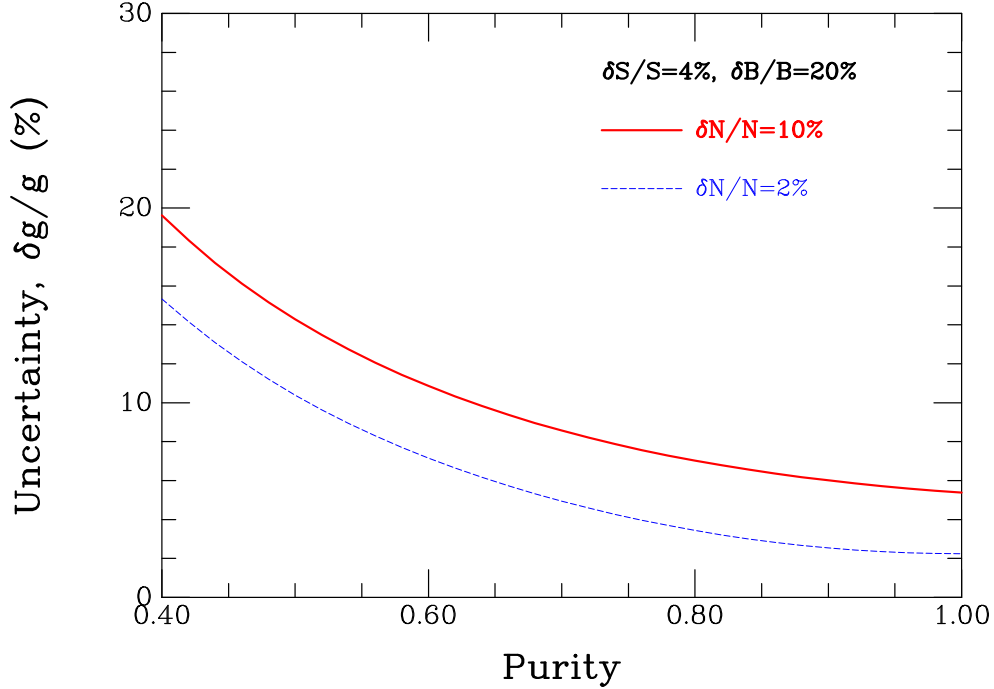,angle=270,width=13cm} \\
\end{center}
\caption{The predicted uncertainty $\delta g/g$ in the coupling of the Higgs 
boson to a pair of $W$ bosons is shown as a function of signal purity 
$P = S/(S+B)$ for expected statistical accuracies $\delta N/N$ of 10\% and 2\% 
The uncertainties in knowledge of the signal $S$ and background $B$ are
assumed to be 4\% and 20\% respectively.}
\label{fig:uncert_lesh}
\end{figure}

Somewhat smaller values of $\delta S/S$ and $\delta B/B$ are chosen in another recent 
investigation of anticipated uncertainties in the couplings~\cite{Duehrssen}.  These 
values are $\delta S/S =4$\% and $\delta B/B = 20$\%.  Although the scope of that 
study is quite different from ours, we may compare our estimates with theirs. In 
Fig.~\ref{fig:uncert_lesh}, we show the uncertainty as a function of purity for these 
new estimates of $\delta S/S$ and $\delta B/B$.  For $P = 0.7$, we now find 
$\delta g/g \sim 9$\% and $\sim 5$\% for the low- and high-luminosity data samples.  This  
new lower value of $\delta g/g$ is similar to that obtained in Ref.~\cite{Duehrssen} at 
comparable luminosity {\footnote {One must bear in mind that the uncertainty discussed 
in Ref.~\cite{Duehrssen} is the uncertainty on $g^2$ and therefore a factor of 2 greater.}}. 

\section{Alternative Definitions of the WBF Sample}
\label{sec:alternatives}

In the previous section, we define the WBF sample by a simple selection 
on the rapidity of one jet in an event in which there is a Higgs boson and two jets 
each carrying $p_T$ greater than a specified minimum value.  Other definitions have 
been used in the literature, and we wish to compare our signal rates and purities with 
those obtained if we  use these alternatives.  We examine the traditional cut on 
rapidity separation between 
the two trigger jets and a cut on the invariant mass of the pair of trigger jets.  

\subsection{Rapidity separation cut}

\begin{figure}[h]
\begin{center} 
\epsfig{file=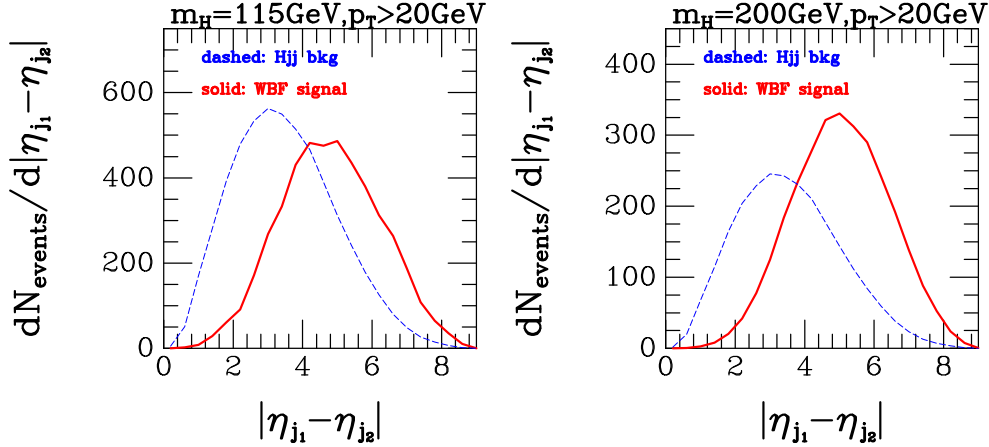,angle=270,width=13cm}
\end{center}
\caption{The difference between the two tagging jet
pseudo-rapidities for a minimum jet $p_T$ of $20$~GeV,
for the two cases $m_H=115$~GeV (left) and $m_H=200$~GeV
(right).
The rates assume an integrated
luminosity of $1~{\rm fb}^{-1}$. The signal (solid) is calculated at
NLO and the background (dashed) at LO.}
\label{fig:absrap_pt_rapgap}
\end{figure}

In Fig.~\ref{fig:absrap_pt_rapgap}, we illustrate the expected event rates as a 
function of the difference in rapidities between the forward and backward tagging 
jets.  There is a clear separation in the locations of the peaks of the WBF 
signal and the background, not unlike that seen in our Fig.~\ref{fig:absrap_pt}.  
Distributions such as these may motivate the choice of a cut on rapidity 
separation, $|\eta_{j1}-\eta_{j2}|>4$, as in 
Refs.~\cite{Kinnunen:1999ak}--\cite{Cavalli:2002vs}, and~\cite{Figy:2003nv}.  
Signal and background rates for WBF events selected in this fashion are shown in 
Table~\ref{tab:rates_rapgap}.  Comparison of Tables~\ref{tab:rates_rapsep}
and ~\ref{tab:rates_rapgap} shows that the signal rate is diminished somewhat and 
that the purity is greater when the rapidity separation selection is made.  As shown 
in Figs.~\ref{fig:uncert} and~\ref{fig:uncert_lesh}, a gain in purity reduces the 
uncertainty in $\delta g/g$.  The quantitative shift from $P = 0.67$ to $P = 0.78$ 
at $M = 115$~GeV and $p_T > 40$~GeV is an improvement of only $3$\% in $\delta g/g$, 
and this reduction is offset somewhat by the loss in statistical accuracy.   

\begin{table}
\begin{center}
\begin{tabular}{|l|c|c|c|} \hline
$p_T$ cut [GeV]      & $20$ & $40$ & $80$ \\ \hline
Signal ($m_H=115$)   & 1297 &  718 &  137 \\ \hline
Bkg                  &  758 &  207 &   38 \\ \hline 
Purity               & 0.63 & 0.78 & 0.78 \\ \hline \hline
Signal ($m_H=200$)   &  911 &  521 &  106 \\ \hline
Bkg                  &  349 &  102 &   20 \\ \hline
Purity               & 0.72 & 0.84 & 0.84 \\ \hline
\end{tabular}
\caption{Event rates for the $Hjj$ WBF signal(NLO) and 
$Hjj$ background(LO), without our WBF definition and instead
with the rapidity separation cut $|\eta_{j_1}-\eta_{j_2}|>4$.  
We assume $1~{\rm fb}^{-1}$ of 
luminosity. The purity and significance are as defined before.}
\label{tab:rates_rapgap}
\end{center}
\end{table}
At lowest order in perturbation theory, one might expect naively that our 
simple rapidity selection and the rapidity separation cut are close to 
identical since there are only two jets in the event, tending to be 
located in opposite hemispheres.  However, the finite rapidity carried by 
the Higgs boson introduces differences.   Our preference for 
the simple rapidity cut is based on a few considerations.  In data 
(and at yet-higher orders in perturbation theory), there will be many jets, 
and the simple specification of events that satisfy Eq.~(\ref{etapeak}) 
will be easier 
to implement. Second, in a high luminosity environment with more than one 
event per beam crossing, a selection on only one jet (in addition to the 
Higgs boson) reduces the chance that jets from different events are used.  
Finally, in our study of NLO event topologies with three jets in the 
final state, we find that a gluon jet, rather than a quark jet is sometimes 
one of the two jets with largest $p_T$. For example, with a jet cut of 
$20$~GeV and $m_H=115$~GeV, a gluon is a tagging jet about $25\%$ of the 
time when we use our definition of WBF events.

\subsection{Invariant mass cut}
\begin{figure}[h]
\begin{center} 
\epsfig{file=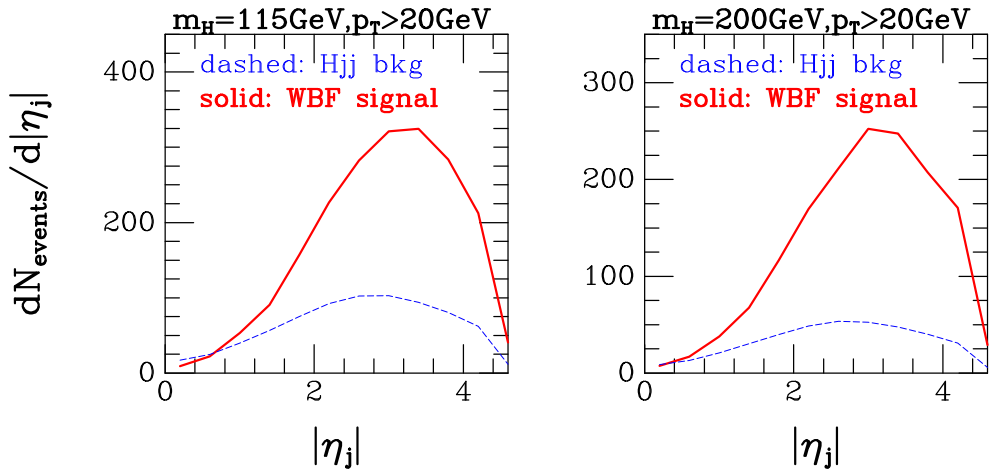,angle=270,width=13cm}
\end{center}
\caption{The tagging jet pseudo-rapidity distribution for a minimum
jet $p_T$ of $20$~GeV, for the two cases $m_H=115$~GeV (left) and $m_H=200$~GeV
(right) -- with a minimum dijet invariant mass, $M_{jj}>800$~GeV.
Each of the two tagging jets in the event is entered in these
plots, with weight one-half, and the rates assume an integrated
luminosity of $1~{\rm fb}^{-1}$. The signal (solid) is calculated at
NLO and the background (dashed) at LO.}
\label{fig:absrap_pt_mcut}
\end{figure}

As an alternative to the rapidity separation cut, one might consider a cut on 
the invariant mass of the two trigger jets.  In Fig.~\ref{fig:absrap_pt_mcut} 
and in Table~\ref{tab:rates_mcut}, we display the the effects of the mass cut 
$M_{jj}>800$~GeV.  Comparison of Figs.~\ref{fig:absrap_pt_mcut} 
and~\ref{fig:absrap_pt} shows a decided improvement in the signal to 
background ratio, an effect that is borne out in the purity numbers shown in 
Tables~\ref{tab:rates_mcut} and~\ref{tab:rates_rapsep}. However, the 
significant gain 
in purity is true only for the smaller values of the $p_T$ cut and is 
accompanied by a substantial loss of signal rate.  Since the smallest 
value of the $p_T$ could be employed only with low-luminosity data samples, 
it is not evident that the price in loss of signal rate is affordable.               
The combination of the mass cut and our simple forward jet cut 
improves purity only slightly and reduces the signal rate further.

\begin{table}
\begin{center}
\begin{tabular}{|l|c|c|c|} \hline
$p_T$ cut [GeV]      & $20$ & $40$ & $80$ \\ \hline
Signal ($m_H=115$)   &  808 &  561 &  158 \\ \hline
Bkg                  &  304 &  183 &   82 \\ \hline 
Purity               & 0.73 & 0.75 & 0.66 \\ \hline \hline
Signal ($m_H=200$)   &  617 &  428 &  121 \\ \hline
Bkg                  &  157 &   95 &   43 \\ \hline
Purity               & 0.80 & 0.82 & 0.74 \\ \hline
\end{tabular}
\caption{Event rates for the $Hjj$ WBF signal(NLO) and 
$Hjj$ background(LO), without our usual WBF definition and instead
with $M_{jj}>800$~GeV. We assume $1~{\rm fb}^{-1}$ of 
luminosity.}
\label{tab:rates_mcut}
\end{center}
\end{table}

Using the invariant mass cut to define the WBF sample, we note that the 
number of events at $m_H = 115$~GeV with a cut on $p_T$ of $20$~GeV is 
very similar to what we obtain with our WBF prescription but with a cut 
on $p_T = 40$~GeV.  The purities are also nearly the same.  This comparison 
would seem to favor our simple prescription: a larger value of the cut on 
$p_T$ is more appropriate in a high rate environment and more effective at 
reducing backgrounds not considered here.  We conclude this section with 
the remark that the alternative prescriptions of the WBF sample in terms of 
either a rapidity separation cut or an invariant mass cut yield some increase 
in the signal purity with respect to our simple cut on the rapidity of one 
jet, but the gain depends on the value of the cut on $p_T$ of the jets and is 
accompanied by some loss of event rate.  As long as hard matrix elements are 
used to generate the $p_T$ distributions of the jets, all three methods yield 
similar event samples.  On the other hand, significant differences seem to 
result if one uses a parton shower method to generate the $H+2$~jet 
background~\cite{Asai:2004ws}.  In our view, the hard matrix element 
approach is a more faithful representation of the momentum distributions of 
jets in the relevant WBF region of large $p_T$.  
  
\section{Estimates of the $Hjj$ NLO corrections}
\label{sec:estimating}

A differential calculation of the $Hjj$ background does not
exist at next-to-leading order. Prior to undertaking such an 
effort, we wish to obtain plausible estimates of the sizes of 
NLO effects on both event rates and kinematic distributions.  We 
present two such estimates in this section. 

\begin{figure}[h]
\begin{center}
\epsfig{file=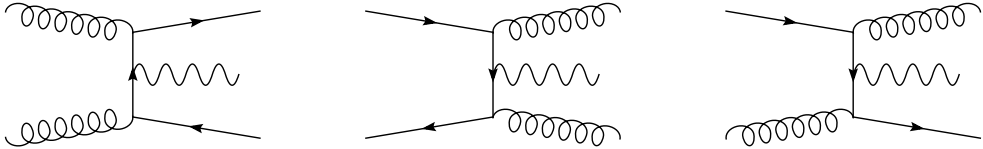,angle=0,width=13cm}
\end{center}
\caption{Representative diagrams for leading order $Z+2$~jet
production.} 
\label{fig:z2jdiags}
\end{figure}

The NLO corrections for QCD production of a $Z$ boson in association 
with two jets are
known~\cite{Campbell:2002tg,Campbell:2003hd}. Representative diagrams
for this process -- seen in Fig.~\ref{fig:z2jdiags} -- can be compared
with those for $H$ plus two jets, shown in Fig.~\ref{fig:h2jdiags}.
Our first estimate of NLO effects for $Hjj$ is based on its similarity 
with $Zjj$, but we acknowledge some important differences.
In the $Hjj$ process the Higgs boson couples only to gluons (via a top 
quark loop), whereas the $Z$ boson couples only to quarks.  This difference 
means that the processes have a different sensitivity to
the parton distribution functions.  Second, the couplings of the scalar 
Higgs boson to the decay products are also different, and the 
angular distributions of the decay products differ in the two cases.
While we do not include a decay branching ratio, our cuts 
require the rapidities of decay products of the produced 
boson to lie between those of the two tagging jets.  Finally, the 
effective $Hgg$ coupling contains a factor of $\alpha_s$ so that the 
$Hjj$ process is formally proportional to $\alpha_s^4$, in contrast with 
$\alpha_s^2$ for $Zjj$.  Although the $Hjj$ process is naively of 
${\cal O}(\alpha_s^4)$, our use an effective coupling for the $Hgg$ vertex 
means that the $Hjj$ process is effectively of the same order, 
${\cal O}(\alpha_s^2)$, as the $Zjj$ process.  Since we are interested only 
in Higgs boson masses that satisfy $m_H<2 m_t$, with transverse momenta 
$p_T^H < m_t$, the effective coupling approach is valid~\cite{DelDuca:2001fn}.

We calculate the $Zjj$ cross section using a variable $Z$-mass, $m_Z=m_H$. 
In this way, we estimate the NLO corrections for a process involving the QCD 
production of a heavy vector boson and two jets. We examine the distribution 
of the tagging jet pseudo-rapidities, as before.  In particular, we are 
interested in whether the shape of this distribution changes significantly 
at NLO.  A NLO effect that would modify the background distribution so that 
it resembles the signal peak, could have a serious effect on the ability to 
select the genuine WBF events.

\begin{figure}[h]
\begin{center}
\epsfig{file=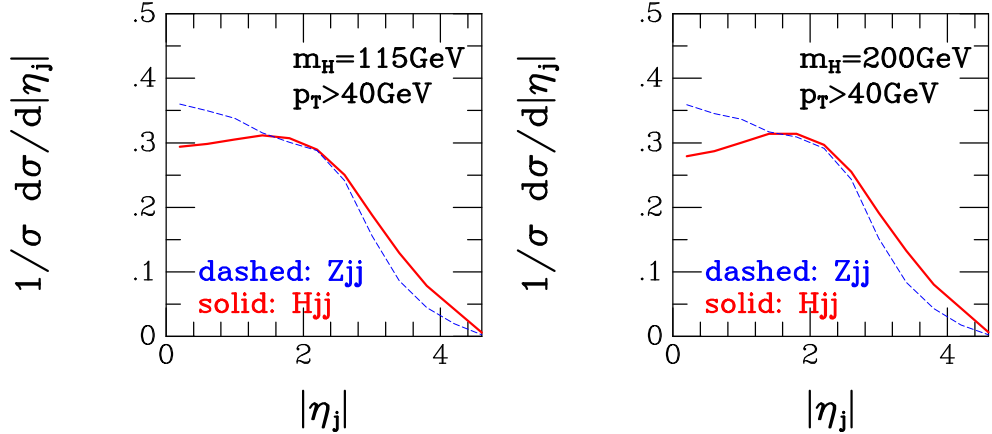,angle=270,width=13cm} \\
\end{center}
\caption{Normalized LO tagging jet pseudo-rapidity distributions in $Hjj$ 
(solid) and $Zjj$ (dashed) events.  The minimum jet $p_T$ cut is $40$~GeV; 
$m_H=115$~GeV (left) and $m_H=200$~GeV (right). For
the $Zjj$ events, the $Z$ mass has been altered to $m_Z=m_H$.}
\label{absrap_norm}
\end{figure}

We show the lowest order distributions for the $Hjj$ and $Zjj$ processes 
in Fig.~\ref{absrap_norm}.
We do not reproduce the plots for more than one value of the $p_T$ cut,
since this cut does not alter the conclusions.  One can see that the 
distributions for the two processes are very similar in shape in the two 
cases, differing in the behavior at low pseudo-rapidities -- where the 
$Zjj$ curve is somewhat higher -- and toward the tail, where the $Hjj$ 
distribution dies off slightly more slowly.  For the purposes of this 
study, the most prominent difference -- that at low values of $|\eta|$ 
-- has little effect since our cuts require that $|\eta_j|>1.6$ for at
least one of the tagging jets. The remaining 
difference in shapes is small, and we conclude that, as a first estimate, 
our use of the $Zjj$ process to approximate $Hjj$ is reasonable.

\begin{figure}[h]
\begin{center}
\epsfig{file=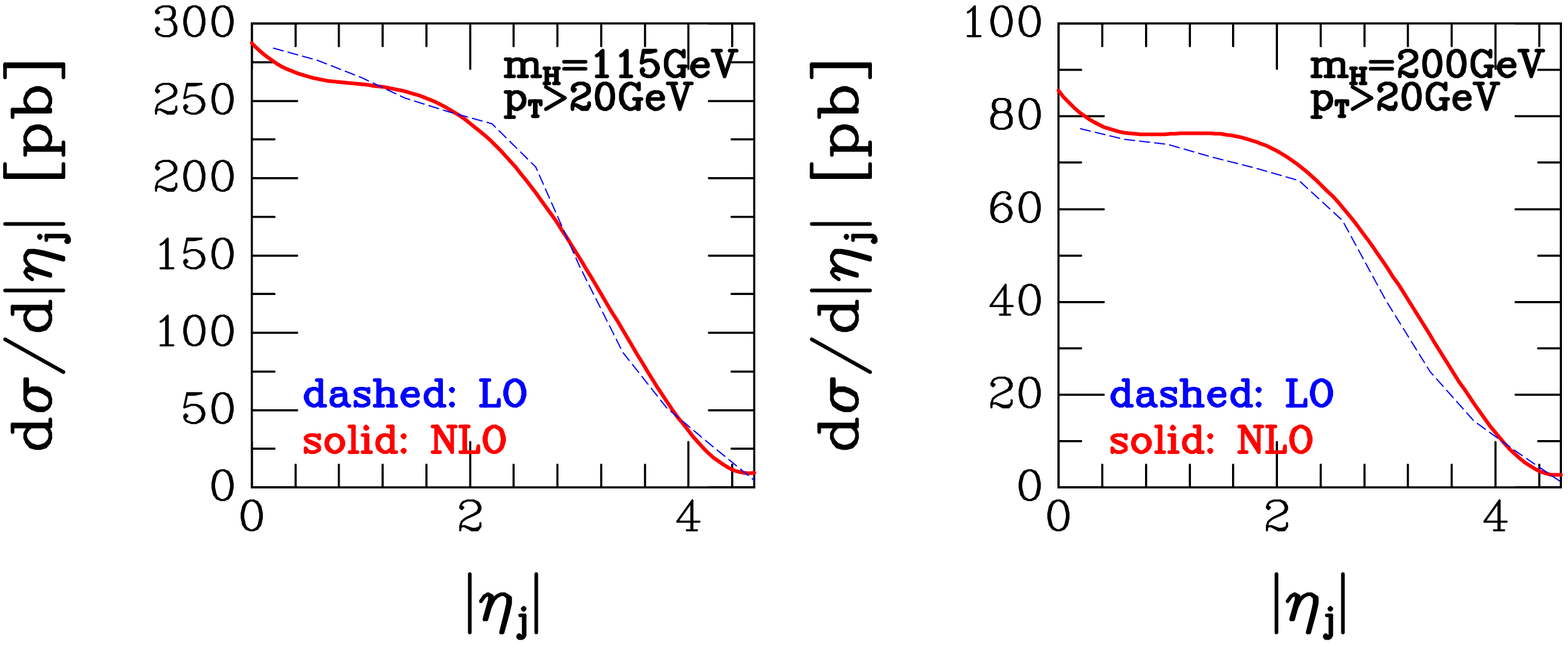,angle=270,width=13cm} \\
\epsfig{file=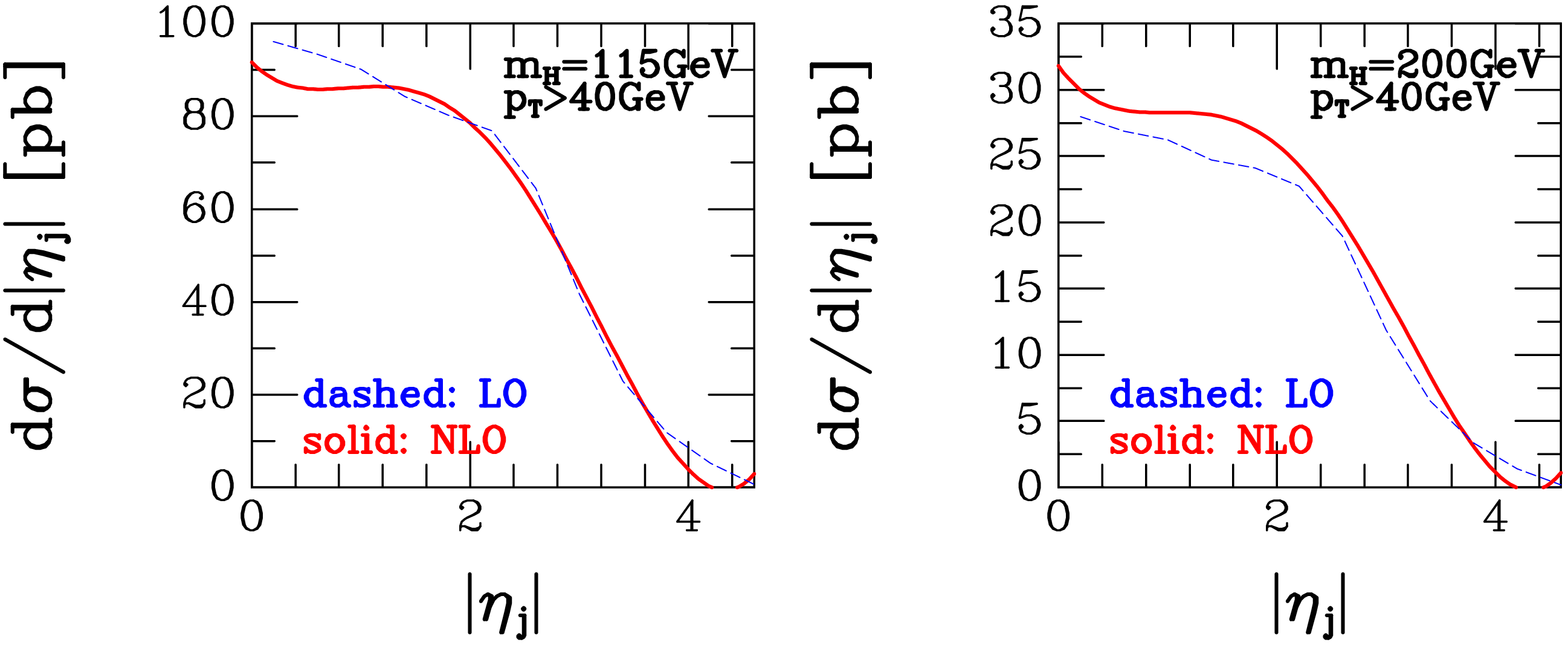,angle=270,width=13cm} \\
\epsfig{file=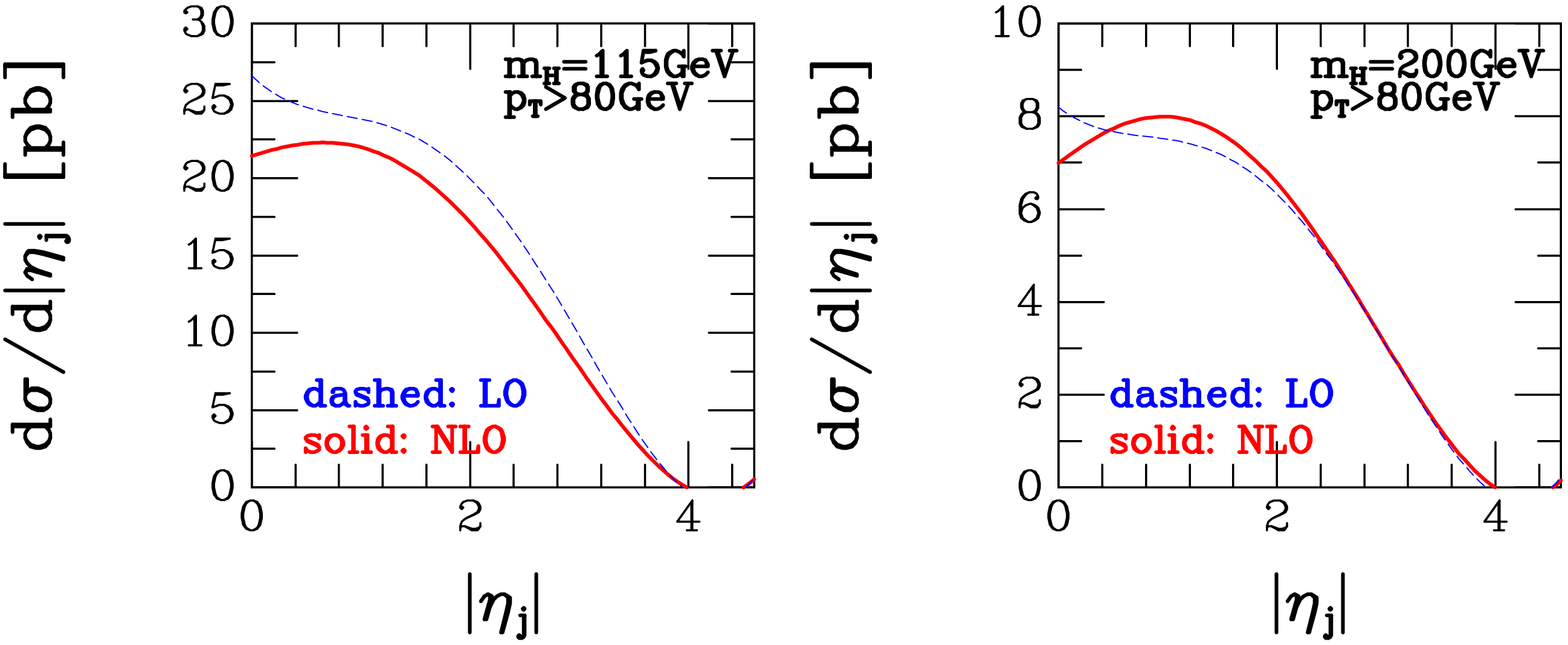,angle=270,width=13cm} \\
\end{center}
\caption{Dependence of the tagging jet pseudo-rapidities in $Zjj$ events
on the minimum jet $p_T$ cut, for the two cases
$m_Z=115$~GeV (left) and $m_Z=200$~GeV (right). In each graph, the LO
(dashed) and NLO (solid) curve is shown.  The scale choice is $\mu = m_H$.}
\label{absrap_pt_zjj}
\end{figure}

For the $Zjj$ process, the lowest order and NLO distributions in the 
tagging jet pseudo-rapidity are shown in
Fig.~\ref{absrap_pt_zjj}.  These are to be compared with the $Hjj$ curves 
in Fig.~\ref{fig:absrap_pt}.
We show curves for two different boson masses, as before, and for a
variety of $p_T$ cuts. The NLO corrections do not appear to alter the shape 
of this distribution significantly.  Moreover, with the scale choice $\mu = m_H$, 
the corrections are small in magnitude, both in the total 
cross section and over the pseudo-rapidity range of interest. The 
change in cross section is shown quantitatively in Table~\ref{tab:kfac}, 
where we show the $K$-factors for this process having applied all 
the WBF cuts.  The corrections vary from $\approx 10\%$ for
$m_Z=200$~GeV and moderate $p_T = 20$, $40$~GeV to $\approx -15\%$ for
$m_Z=115$~GeV and high $p_T =80$~GeV.

\begin{table}
\begin{center}
\begin{tabular}{|l|c|c|c|} \hline
$p_T$ cut [GeV]      & $20$ & $40$ & $80$ \\ \hline
$m_H=115$            & 1.00 &  0.99 & 0.84 \\ \hline
$m_H=200$            & 1.12 &  1.11 & 1.02 \\ \hline
\end{tabular}
\caption{The $K$-factors, as defined by Eq.~\ref{eq:kfac},
for $Zjj$ production with our WBF cuts. The $Z$ mass has been
altered to take on the two relevant Higgs boson mass values, $m_Z=m_H$, 
and the scale choice is $\mu = m_H$.}
\label{tab:kfac}
\end{center}
\end{table}

In the definition of the $K$-factor in this paper, different parton 
distribution functions (PDF's) are used in the numerator and denominator, 
LO expressions in the denominator and NLO expressions in the numerator.    
Correspondingly, different values of $\alpha_s$ are used in the numerator 
and denominator:
\begin{equation}
K=\frac{\sigma^{NLO}({\rm CTEQ6M}; \alpha_s^{NLO}(\mu))}
       {\sigma^{LO} ({\rm CTEQ6L1}; \alpha_s^{LO}(\mu))}
\label{eq:kfac}
\end{equation}
That the $K$-factors are close to unity for the $Zjj$ process results 
from a compensation between the change in PDF from LO to NLO and the change 
in $\alpha_s(\mu)$, plus the effects of the additional processes at NLO.  
In $\alpha_s(\mu)$, the 
net change, after reduction in the value of $\alpha_s(M_Z)$ and the altered 
evolution, tends to decrease the cross section from LO to NLO, by a
factor
$$
\left[\frac{\alpha_s^{NLO}(\mu)}{\alpha_s^{LO}(\mu)}\right]^2=0.83,
$$
for both cases ($\mu=115$, $200$~GeV).
One must be careful to apply this $K$-factor consistently only to a lowest 
order calculation with the same PDF set and treatment of $\alpha_s$.
Since we use CTEQ6L1 in the background calculation of
Table~\ref{tab:rates_rapsep}, it is straightforward to incorporate
these $K$-factors in order to obtain the new background estimates in
Table~\ref{tab:rates_rapsep_nlo}, for an assumed $1~{\rm fb}^{-1}$
of integrated luminosity.  For the estimate of NLO 
corrections to the background presented in this section, with 
$\mu = m_H$, the table shows that the 
purity starts at about 50\% at $m_H = 115$~GeV, if the cut on $p_T$ is 
$20$~GeV, and grows to about 70\% when the $p_T$ cut is $40$~GeV or larger.  
Slightly larger values are obtained at $m_H = 200$~GeV. 
\begin{table}
\begin{center}
\begin{tabular}{|l|c|c|c|} \hline
$p_T$ cut [GeV]      & $20$ & $40$ & $80$ \\ \hline
Signal ($m_H=115$)   & 1374 &  789 &  166 \\ \hline
Bkg                  & 1196 &  378 &   77 \\ \hline
Purity               & 0.53 & 0.68 & 0.68 \\ \hline \hline
Signal ($m_H=200$)   &  928 &  545 &  121 \\ \hline
Bkg                  &  598 &  199 &   47 \\ \hline
Purity               & 0.61 & 0.73 & 0.72 \\ \hline
\end{tabular}
\caption{Event rates for the $Hjj$ WBF signal(NLO) and 
$Hjj$ background(estimated NLO), for an assumed $1~{\rm fb}^{-1}$ of 
luminosity.  No branching ratio is included for the Higgs boson decay.}
\label{tab:rates_rapsep_nlo}
\end{center}
\end{table}
 
\begin{figure}[h]
\begin{center} 
\epsfig{file=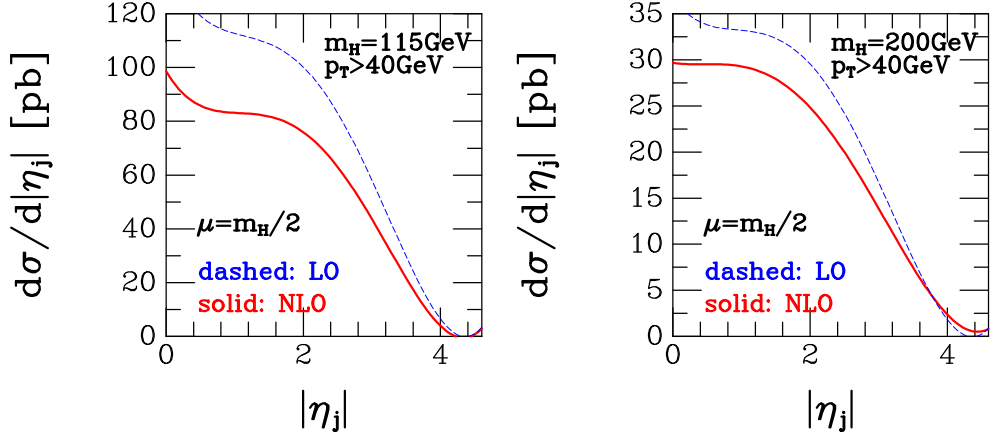,angle=270,width=13cm} \\
\end{center}
\caption{Tagging jet pseudo-rapidity distributions in $Zjj$ events, calculated
with a smaller renormalization and factorization scale, $\mu=m_H/2$.
In each graph, the LO (dashed) and NLO (solid) curve is shown.}
\label{absrap_scale_zjj}
\end{figure}

\begin{figure}[h]
\begin{center} 
\epsfig{file=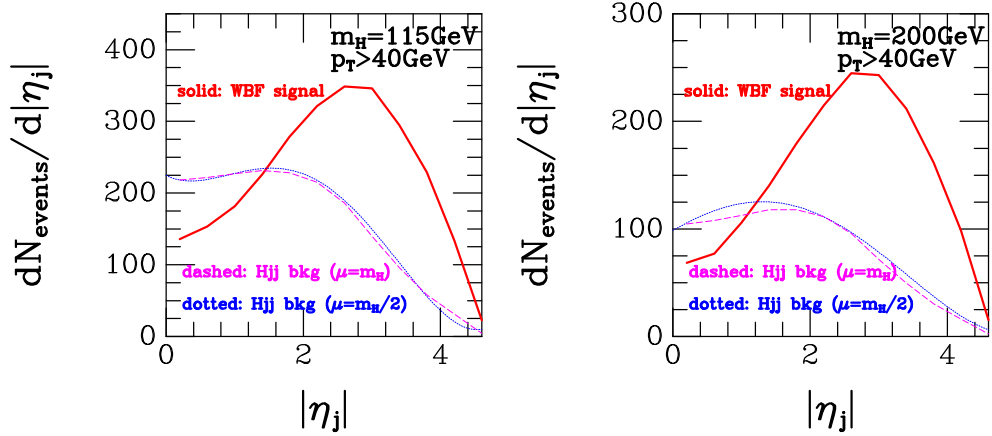,angle=270,width=13cm} \\
\end{center}
\caption{Estimated tagging jet pseudo-rapidity distributions in $Hjj$ events 
at NLO,
calculated with a smaller renormalization and factorization scale,
$\mu=m_H/2$. Also shown (lower curve) is the estimated NLO $Hjj$ result
with $\mu=m_H$.}
\label{absrap_scale_nlo}
\end{figure}

We show in Fig.~\ref{absrap_scale_zjj} the effect of a lower scale choice 
$\mu = m_H/2$ on the $Zjj$ process.  In contrast to Fig.~\ref{absrap_pt_zjj}, 
the NLO corrections are now substantial, negative, and not constant as the 
pseudo-rapidity changes.
For this lower scale choice, we now apply the $K$-factors point-by-point to 
the lowest order $Hjj$ background distribution in order to estimate the new 
NLO result. The result of this exercise is shown in
Fig.~\ref{absrap_scale_nlo}, along with the estimated NLO
result for $\mu=m_H$. 
This figure shows that our estimate for the NLO $Hjj$
background cross section is affected very little if we choose a smaller scale 
such as $\mu=m_H/2$.  Reducing the scale, we find that the LO background 
is increased substantially.  However, the $K$-factor decreases in such a 
way as to restore the size of background at NLO.  

\subsection{Background estimate with a mass cut}

\begin{figure}[h]
\begin{center}
\epsfig{file=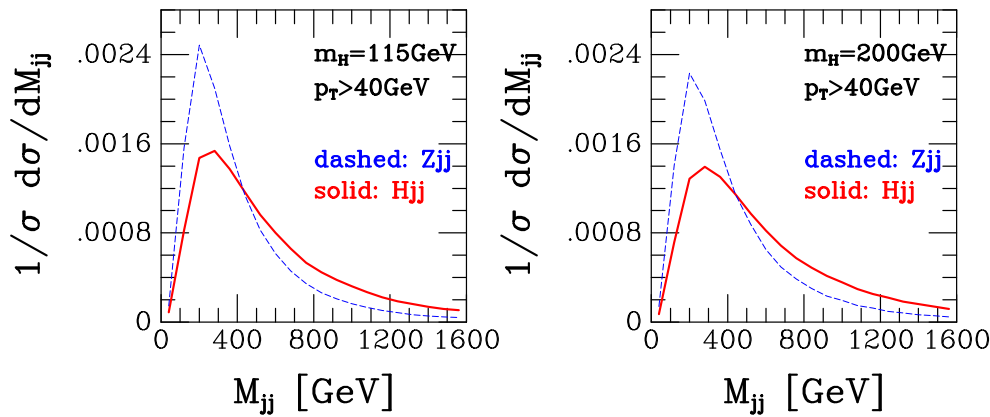,angle=270,width=13cm} \\
\end{center}
\caption{Normalized LO tagging jet invariant mass distributions in $Hjj$ 
(solid) and $Zjj$ (dashed) events.  The minimum jet $p_T$ cut is $40$~GeV; 
$m_H=115$~GeV (left) and $m_H=200$~GeV (right). For
the $Zjj$ events, the $Z$ mass has been altered to $m_Z=m_H$.}
\label{mjj_norm}
\end{figure}

In a similar spirit to the above study, here we present the estimate
of the NLO corrections to the $Hjj$ background when the definition
of the WBF event sample involves only an invariant mass cut on the two
tagging jets, $M_{jj}>800$~GeV.  In Fig.~\ref{mjj_norm}, we show the 
lowest order
distribution of the invariant mass of the two tagging jets, for the
processes $Hjj$ and $Zjj$ and a jet $p_T$ cut of $40$~GeV.
Each curve is normalized by its own integrated cross section.
Although the two distributions differ in shape considerably over such
a wide range of $M_{jj}$, the two curves show similar behavior above 
$M_{jj}>800$~GeV.  We conclude again that the
$Zjj$ process will yield a reasonable estimate of the $Hjj$ process in 
the region of phase space of interest for WBF studies.

\begin{figure}[h]
\begin{center}
\epsfig{file=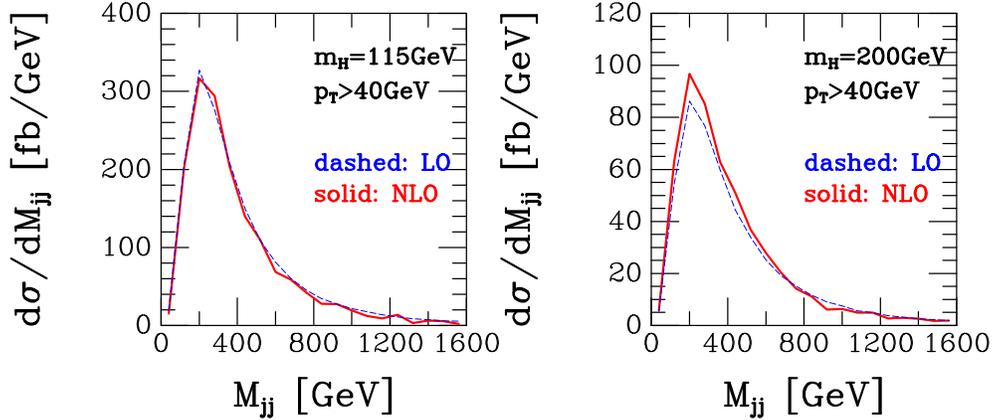,angle=270,width=13cm} \\
\end{center}
\caption{Tagging jet invariant mass distribution in $Zjj$ events
with a jet $p_T$ cut of $40$~GeV, for the two cases
$m_Z=115$~GeV (left) and $m_Z=200$~GeV (right). The LO
(dashed) and NLO (solid) curves are shown, for a scale choice of
$\mu = m_H$.}
\label{mjj_zjj}
\end{figure}

The effects of the NLO corrections for the $Zjj$ process are shown in
Fig.~\ref{mjj_zjj}.  They are small over the entire mass range, and they 
do not change the shape of the distribution. The net effect is
summarized in Table~\ref{tab:kfac_mjj}, where we show the $K$-factors for
each $p_T$ cut and Higgs mass for the alternative definition of the 
WBF sample, $M_{jj}>800$~GeV.
\begin{table}
\begin{center}
\begin{tabular}{|l|c|c|c|} \hline
$p_T$ cut [GeV]      & $20$ & $40$ & $80$ \\ \hline
$m_H=115$            & 0.89 &  0.80 & 0.78 \\ \hline
$m_H=200$            & 0.93 &  0.93 & 0.89 \\ \hline
\end{tabular}
\caption{The $K$-factors, as defined by Eq.~(\ref{eq:kfac}),
for $Zjj$ production with an alternative definition of the
WBF cuts, $M_{jj}>800$~GeV. The $Z$ mass has been
altered to take on the two relevant Higgs boson mass values, $m_Z=m_H$.}
\label{tab:kfac_mjj}
\end{center}
\end{table}
The corrections are universally negative, ranging from $\approx -10\%$
for $m_Z=200$~GeV to $\approx -20\%$ for
$m_Z=115$~GeV and the higher $p_T$ cuts of $40$,$80$~GeV.

Using these $K$-factors, we can update the earlier leading order
results of Table~\ref{tab:rates_mcut}. Since the $K$-factors are close
to unity, the effect of the NLO corrections is small, as illustrated
in Table~\ref{tab:rates_mcut_nlo}.
\begin{table}
\begin{center}
\begin{tabular}{|l|c|c|c|} \hline
$p_T$ cut [GeV]      & $20$ & $40$ & $80$ \\ \hline
Signal ($m_H=115$)   &  808 &  561 &  158 \\ \hline
Bkg                  &  271 &  146 &   64 \\ \hline 
Purity               & 0.75 & 0.79 & 0.71 \\ \hline \hline
Signal ($m_H=200$)   &  617 &  428 &  121 \\ \hline
Bkg                  &  146 &   88 &   38 \\ \hline
Purity               & 0.81 & 0.83 & 0.76 \\ \hline
\end{tabular}
\caption{Event rates for the $Hjj$ WBF signal(NLO) and 
$Hjj$ background(estimated NLO), without our usual WBF definition and instead
with $M_{jj}>800$~GeV. We have assumed $1~{\rm fb}^{-1}$ of 
integrated luminosity.}
\label{tab:rates_mcut_nlo}
\end{center}
\end{table}
With NLO effects included, we conclude as at LO, that the cut on the jet-pair 
invariant mass improves the signal purity significantly 
with respect to our simple rapidity prescription only for the smallest of 
the cuts on $p_T$ ({\it c.f.} results Table~\ref{tab:rates_rapsep_nlo}), 
but at a cost in the signal rate at all $p_T$.  
 
\subsection{Second Estimate of NLO Corrections to the Background}
\label{sec:est2}

A different and larger estimate of the NLO corrections to the $Hjj$ 
background may be obtained if we begin with results for 
inclusive Higgs boson production processes.  Corrections to the 
fully inclusive cross section are known both to 
NNLO~\cite{Catani:2001ic}--\cite{Ravindran:2003um} and with all-orders soft-gluon 
resummation included~\cite{Berger:2002ut}--\cite{Berger:2003pd}.
In addition, the NLO corrections to the $H+j$ process are 
published~\cite{Ravindran:2002dc}.  By starting from the inclusive 
process or the $H+j$ process, we begin with a process that is 
dominated by $gg$ scattering, unlike the $Zjj$ case.    

With the same definition of the $K$-factor used in this paper, and 
the same choice of renormalization/factorization scale, the result for 
the fully inclusive Higgs boson cross section is 
$$
K(H+X)=1.7 - 1.8.  
$$
The increase at NLO is slightly larger at $m_H=200$~GeV than
at $m_H=115$~GeV.  For the semi-inclusive $H+j$ process, where an 
additional jet of $p_T>20$--$80$~GeV is specified in the final state,
the NLO corrections correspond to,
$$
K(H+j+X) \approx 1.3 - 1.5.  
$$
In this case the PDF sets used were similar but not identical
(CTEQ4), and this result is quoted for $m_H=120$~GeV.

We now try to apply these $K$ factors, obtained in situations 
in which either no jet or at most one jet is required in the final 
state, to the case of interest here where two jets are required, each 
with a minimum value of $p_T$, and with the other WBF cuts.  In the 
case of $Z$~$+$~jets, the $K$-factors are $1.14$, $1.16$, and $0.90$ 
for $0$, $1$, and $2$ jets, respectively~\cite{Campbell:2003hd}, with 
the same PDF's and definition of $K$-factors used here.  In other 
words, the $K$-factor is approximately independent of the number of 
specified jets.  In Table~\ref{tab:kfac}, 
$K$-factors are shown for the $Zjj$ process with WBF cuts applied.  
They are close to unity.  In Ref.~\cite{Campbell:2003hd}, the $K$-factor 
is obtained for the $Z+2$~jet process with a jet cut of $p_T > 20$~GeV, 
but without the WBF cuts.  It is also close to unity, $K=0.9$. This value 
is similar enough to the numbers shown in Table~\ref{tab:kfac} to justify 
the assumption that the overall inclusive $K$-factor is not very different 
from the one that should be applied for WBF events.  Adopting this line 
of argument, we suggest that a conservative estimate of the $K$-factor 
for $Hjj$ production in the WBF region is a $K$-factor from the more 
inclusive processes above, 
$$
K(H+jj+X) \approx \frac{K(H+X)+K(H+j+X)}{2} = 1.6.
$$
The new estimates of NLO corrections to the background are presented 
in Table~\ref{tab:rates_rapsep_nlo_est2}.  The signal purity starts at 
about 40\% when $m_H = 115$~GeV, 
if the cut on $p_T$ is $20$~GeV, and grows to about 60\% when the $p_T$ cut 
is $40$~GeV or larger.  Slightly larger values are obtained at $m_H = 200$~GeV.  

With the larger background estimate of this subsection, the signal purity is 
lower than before ({\it c.f.} Table~\ref{tab:rates_rapsep_nlo_est2} {\it vs} 
Table~\ref{tab:rates_rapsep_nlo}).  If the NLO enhancement of the background is 
as great as a factor 1.6, the value of $P \sim 0.4$ presented in 
Table~\ref{tab:rates_rapsep_nlo_est2} suggests that investigations with a low 
$p_T$ cut will be ineffective.  According to the results in 
Fig.~\ref{fig:uncert}, 
the uncertainty in $\delta g/g$ is quite large when $P < 0.5$.  With the cut 
$p_T \ge 40$~GeV, the corresponding purities of $P \sim 0.6$ may still permit 
determinations of $\delta g/g \sim 10$\%.  A fully 
differential NLO calculation of the background process for $H+2$~jets is 
definitely needed to establish both the size of the background and the 
theoretical uncertainty $\delta B/B$.  

\begin{table}
\begin{center}
\begin{tabular}{|l|c|c|c|} \hline
$p_T$ cut [GeV]      & $20$ & $40$ & $80$ \\ \hline
Signal ($m_H=115$)   & 1374 &  789 &  166 \\ \hline
Bkg                  & 1914 &  611 &  123 \\ \hline 
Purity               & 0.42 & 0.56 & 0.57 \\ \hline \hline
Signal ($m_H=200$)   &  928 &  545 &  121 \\ \hline
Bkg                  &  854 &  286 &   74 \\ \hline
Purity               & 0.52 & 0.67 & 0.62 \\ \hline
\end{tabular}
\caption{Event rates for the $Hjj$ WBF signal(NLO) and 
$Hjj$ background(estimated NLO, according to the procedure in
Section~\ref{sec:est2}), for an assumed $1~{\rm fb}^{-1}$ of 
integrated luminosity.}
\label{tab:rates_rapsep_nlo_est2}
\end{center}
\end{table}

\section{Summary and Conclusions}

\label{sec:conclusions}

In this paper, we investigate the weak boson fusion process for production of 
the neutral Higgs boson $H$ at the LHC.  We are interested in estimating the 
accuracy with which the Higgs boson coupling to weak bosons may be determined 
from data.  An important and, in fact, controlling aspect is the extent to 
which events produced by the WBF subprocess may be separated from events in  
which a Higgs boson is produced by other mechanisms.  A hallmark of the WBF 
subprocess is that the Higgs boson is accompanied in the final state by two
jets that carry large transverse momentum $p_T$ and relatively large rapidity.  
However, purely strong interactions subprocesses also produce Higgs bosons 
accompanied by two jets.  To extract the couplings reliably, a good understanding 
is required of both the production and the background processes.  We use
hard QCD matrix elements in order to represent the signal and the $H+2$~jet 
background reliably.  We provide an independent calculation that verifies the 
fully differential next-to-leading order QCD corrections to the WBF 
signal process of Ref.~\cite{Figy:2003nv}, and we examine in more detail the 
effects of the WBF selection cuts on these NLO QCD corrections.  We use 
leading order perturbative QCD expressions for the background $H+2$~jet matrix 
elements since NLO results are not yet available in fully differential 
form.  We also provide two estimates of the NLO enhancement of the QCD $H+2$~jet 
background process.  Our calculations are fully differential at the 
partonic level.  

Among our goals in this study are to evaluate the effectiveness of different 
prescriptions for defining the WBF sample and to estimate the expected 
WBF signal purity $P$, by which we mean the fraction of real Higgs boson events 
produced by weak boson fusion.  

Various prescriptions are used in the literature to define the WBF sample, cuts 
that enhance the WBF fraction of the cross section by exploiting the special 
transverse momentum and rapidity characteristics of WBF events.  Our investigations 
lead us to propose a new, somewhat simpler definition in terms of a {\it cut on the 
rapidity of one of the final state jets}, as defined by Eq.~(\ref{etapeak}).  We 
compare the effects on both event 
rates and signal-to-background ratio of our proposed method for defining WBF events 
with two other popular methods found in the literature: a selection on the difference 
in rapidities between two tagging jets in the final state, and a selection on the 
invariant mass of a pair of tagging jets.  In a low-luminosity environment where 
one may be tempted to use a relatively low cut on transverse momentum to select the 
trigger jets, the conventional alternatives provide better signal purity but at a cost 
in signal rate.  Once the cut is raised, our definition does essentially as well in 
signal purity while preserving more of the signal.  

We find that purities of 60\% to 70\% can be expected if a selection of $p_T \ge 40$~GeV 
is made on the tagging jets.  We derive an expression for the expected uncertainty on 
the effective Higgs-boson-to-weak-boson coupling strength $g$ in terms of $P$, the 
expected statistical accuracy of the LHC experiments, and the theoretical uncertainties 
on the signal and the background processes.  We estimate that an accuracy of 
$\delta g/g \sim 10$\% may be achievable after $\sim 200$~fb$^{-1}$ of integrated 
luminosity is accumulated at the LHC.  On a cautionary note, however, we recall that 
our WBF signal purity and our uncertainties are obtained in a very well controlled 
situation in which there is an identified Higgs boson in a sample of $H + 2$~jet events 
produced by both the WBF mechanism and the QCD background processes.  In an experimental 
context, there will be additional sources of background from final states that mimic a 
Higgs boson.  The effects of these additional backgrounds presumably only increase the 
expected uncertainties on the couplings.   

The theoretical uncertainties on the signal $S$, and on both the size and uncertainty 
of the background $B$ dominate the uncertainty in $g$.  Current estimates of $\delta S/S$ 
are in the $5$\% range, and, since differential NLO calculations exist, this 
uncertainty is controlled by uncertainties in the parton densities and by the residual 
renormalization and factorization scale dependence.  In order to reduce the estimated 
uncertainty in $g$, the next major step would appear to be a fully differential NLO 
calculation 
of the background applicable in the region of interest for WBF investigations.  
  
\section*{Acknowledgments}

\indent\indent  We thank T.~LeCompte for valuable discussions and R.~K.~Ellis for
collaboration in the early stages of this work.  The authors 
gratefully acknowledge the Kavli Institute for Theoretical Physics, Santa Barbara 
for hospitality and
support during the completion of the manuscript. This work was
supported by the U.~S.~Department of Energy under contract
No.~W-31-109-ENG-38 and in part by the National Science Foundation
under Grant No. PHY99-07949.

\end{document}